\documentclass{emulateapj}
\usepackage{latexsym}
\usepackage{dcolumn}
\usepackage{bm}
\input{epsf}
\newcommand{\be}{\begin{equation}}
\newcommand{\ee}{\end{equation}}
\newcommand{\ba}{\begin{eqnarray}}
\newcommand{\ea}{\end{eqnarray}}
\newcommand{\no}{\nonumber}
\newcommand{\bi}{\begin{itemize}}
\newcommand{\ei}{\end{itemize}}
\newcommand{\bfi}{\begin{figure}[t]
\epsfxsize=9cm
\epsffile}
\newcommand{\efi}{\end{figure}}
\newcommand{\apjl}{ApJ}
\newcommand{\apjs}{ApJS}
\newcommand{\mnras}{MNRAS}
\newcommand{\physrep}{Physics Reports}
\newcommand{\araa}{Annual Review of Astronomy \& Astrophysics}
\newcommand{\aj}{Astronomical Journal}

\begin{document}
\title{Self-calibration of gravitational shear-galaxy intrinsic ellipticity
  correlation  in weak  lensing surveys}   
\author{Pengjie Zhang}
\affil{Key Laboratory for Research in Galaxies and Cosmology, Shanghai
  Astronomical Observatory, Nandan Road 80, Shanghai, 200030,
  China}
\email{pjzhang@shao.ac.cn}
\begin{abstract}
The galaxy intrinsic alignment is a severe
challenge to  precision cosmic shear measurement.  We propose to self-calibrate
the induced gravitational shear-galaxy intrinsic ellipticity correlation
(the GI  correlation, \citealt{Hirata04b}) in weak  lensing surveys with
photometric redshift measurement. (1) We  propose a method to extract the intrinsic  
ellipticity-galaxy density cross correlation (I-g) from the galaxy ellipticity-density
measurement in the same redshift bin. (2) We also find a  generic scaling relation
to convert  the extracted I-g correlation to the demanded GI correlation.  We
perform concept study under simplified 
conditions and  demonstrate  its capability to  significantly reduce the GI
contamination.  We discuss the impact of various complexities on the two key
ingredients of the self-calibration technique,  namely the method to extract
the I-g correlation and  the scaling relation between the I-g and the GI
correlation.  We expect none of them is likely able to completely invalidate
the proposed self-calibration technique.  
\end{abstract}
\maketitle
\section{Introduction}
Weak gravitational lensing is one of the most powerful probes of
the dark universe \citep{Refregier03,DETF,Munshi08,Hoekstra08}. It is rich in
physics and contains tremendous information on dark matter,  dark energy and
the nature of gravity at large scales. Its modeling is relatively clean. At
 the  multipole $\ell< 2000$,  
gravity is the dominant  force shaping the weak lensing power spectrum while
complicated gas physics only affects the lensing power spectrum at less than $
1\%$ level \citep{White04,Zhan04,Jing06,Rudd08}. This  makes the weak lensing
precision modeling feasible, through high 
precision simulations. On the other hand, weak lensing has been
measured with high significance. The most sophisticated method so far  is to
measure the cosmic shear,  lensing induced galaxy shape distortion. After the first
detections in the year 2000 \citep{Bacon00,Kaiser00,VanWaerbeke00,Wittman00},
data  quality has been  improved dramatically (e.g. \citealt{Fu08}). 

However, it is still challenging to perform precision lensing measurement. A
potentially severe problem is the non-random galaxy intrinsic
ellipticity (intrinsic alignment). Its existence is supported by many
evidences. Angular momentum of dark matter halos is  
influenced by the tidal force of the large scale structure.  More importantly,
mass accretion is more preferentially along the filaments
(e.g. \citealt{Zhangyc09}). This causes   the dark   matter halo ellipticities
(and also halo orientations) to be correlated over tens of Mpc
\citep{Croft00,Heavens00,Lee00,Catelan01,Crittenden01,Jing02,Zhangyc09}. Although  
galaxies are not perfectly aligned with the halo   
angular momentum vector, with the halo shape, or with each other
\citep{Heymans04,Yang06,Kang07,Okumura08,Wang08}, the  resulting correlations in
galaxy spin  \citep{Pen00} and ellipticity
\citep{Brown02,Mackey02,Heymans04,Hirata04a,Lee07,Okumura08,Schneider09} are  still  measurable and 
contaminate cosmic shear measurement, especially for elliptical galaxies. 

The intrinsic alignment biases the cosmic shear measurement. It introduces an
intrinsic ellipticity-intrinsic ellipticity correlation (the II
correlation). Since this correlation only exists at small physical separation
and hence small line-of-sight separation, it can
be effectively eliminated by the lensing tomography  technique
\citep{King02,King03,Heymans03}, with moderate loss of cosmological
information \citep{Takada04}

However, as pointed out by \citet{Hirata04b} and then confirmed by observations
\citep{Mandelbaum06,Hirata07,Okumura09}, the galaxy intrinsic ellipticity is also
correlated with the large scale structure and thus induces the gravitational
(cosmic) shear-intrinsic ellipticity correlation (the GI
correlation). This contamination could bias the cosmic shear measurement at $10\%$
level and dark energy constraints at more significant level
\citep{Bridle07}. For this reason, there are intensive efforts to correct the GI
contamination  
(e.g. \citealt{Hirata04b,Heymans06,Joachimi08,Joachimi09,Okumura09,
  Joachimi09b,Shi10}).

This paper proposed a new method to alleviate this problem. It
is known that weak lensing surveys contains information other than the
ellipticity-ellipticity correlation (e.g. \cite{Hu04,Bernstein09}).  These
information not only helps improve cosmological constraints, but also helps
reduce errors in lensing measurements, such as photometric redshift errors
\citep{Schneider06,Zhang09}.  We show that these extra information allows for
a promising self-calibration of the GI contamination. The self-calibration technique we
propose relies on no assumptions on the intrinsic ellipticity nor expensive
spectroscopic redshift measurements.\footnote{We still need sufficiently
  accurate photo-z, which controls the I-g measurement error \S \ref{subsec:Ig} and
  the accuracy of the scaling relation ( \S \ref{subsec:systematic} ). Since photo-z
  algorithm relies on calibration against spectroscopic samples (not
  necessarily of the same survey), in this
  sense, our self-calibration does rely on spectroscopic 
  redshift measurements. }  It is thus applicable to ongoing or proposed
lensing surveys like CFHTLS, DES, Euclid, LSST and SNAP/JDEM.  Under simplified
conditions, we estimate the performance of this technique and its robustness
against various sources of error. We find that it has the potential to  reduce
the GI contamination significantly. 

The paper is organized as follows. In \S \ref{sec:GI}, we explain the
basic idea of the self-calibration technique. We investigate the
self-calibration performance  in \S \ref{sec:error}, discuss
extra sources of error in \S \ref{sec:moreerrors} and 
summarize in \S \ref{sec:summary}. We leave some technical details in the
appendices \S \ref{sec:scaling}, \ref{sec:Q} \& \ref{sec:appendixC}. 

\section{A simplified version of the GI self-calibration}
\label{sec:GI}
In the section, we present the basic idea of the self-calibration
technique. To highlight the key ingredients of this technique, we adopt a
simplified picture and neglect many complexities until \S \ref{sec:error} \&
\S \ref{sec:moreerrors}. 

\subsection{The information budget in weak lensing surveys}
\label{subsec:information}
In lensing surveys, we have several pieces of information (refer to
\citet{Bernstein09} for a comprehensive review). What
relevant for the self-calibration technique is the shape, angular position and
photometric redshift of each galaxy. Only galaxies sufficiently large
  and bright are suitable for cosmic shear measurement. To avoid possible
  sample bias,  we will restrict the discussion to  this sub-sample. We split
  galaxies into a set of photo-z bins according to their photo-z $z^P$. The
  $i$-th bin has $\bar{z}_i-\Delta 
  z_i/2\leq z^P\leq \bar{z}_i+\Delta z_i/2$.  Our convention is that, if
  $i<j$, then $\bar{z}_i<\bar{z}_j$.  $\bar{n}^P_i(z^P)$ and $\bar{n}_i(z)$ are
  the mean galaxy distribution  over the $i$-th redshift bin, as a function of
  photo-z $z^P$ and true redshift  $z$ respectively.  The two are related  by
  the photo-z probability distribution function (PDF) $p(z|z^P)$. 
Intrinsic fluctuations in the 3D galaxy distribution, $\delta_g$, cause
fluctuations in the  galaxy surface density of a photo-z bin,
 $\delta_\Sigma$. This  is one piece of information in weak lensing surveys. 

The galaxy shape, expressed in the term of ellipticity, measures the cosmic
shear $\gamma_{1,2}$ induced by gravitational lensing.  For each galaxy, this
signal is overwhelmed  by galaxy intrinsic ellipticity. If galaxy intrinsic
ellipticity  is randomly distributed, it only causes shot noise in the cosmic
 shear   measurement, which is straightforward to correct. For this reason, we
will not explicitly show this term in relevant equations, although we do take
it into account in the error estimation. However, gravitational tidal force
induces correlations in ellipticities of close galaxy pairs. For 
this, the measured shear $\gamma^s_{1,2}=\gamma_{1,2}+I_{1,2}$, where $I$
denotes   the correlated part of the galaxy intrinsic ellipticity and $1,2$
denote the two components of the cosmic shear and the intrinsic
alignment. $\gamma_{1,2}$ describe the shape distortion along the x-y axis and
the direction of $45^{\circ}$ away, respectively. $\gamma$ is  equivalent to the
lensing convergence  
$\kappa$ (in Fourier space, this relation is local). Thus we will work on
$\kappa$ instead of $\gamma_{1,2}$. From the 
measured $\gamma^s$, we obtain  $\kappa^s=\kappa+I$. Here, $I$ is the E mode
of $I_{1,2}$, analogous to $\kappa$, which is  the E mode of $\gamma_{1,2}$
\citep{Stebbins96,Crittenden02}.   Although cosmic shear, to a good
approximation, does not have B-mode,  the intrinsic alignment can have
non-vanishing B-mode (e.g. \citealt{Hirata04b,Schneider09}). This piece of information is useful to diagnose and
hence calibrate the intrinsic alignment. However in the current paper, we will focus
on the E-mode. 

$\kappa$ is the projection of the matter over-density  along the line of sight. For a flat
universe and under the Born approximation, the  lensing convergence $\kappa$
of a galaxy (source) at redshift 
$z_G$ and direction $\hat{\theta}$ \citep{Refregier03} is  
\ba
\label{eqn:kappa}
\kappa(\hat{\theta})=\int_0^{\chi_G} \delta_m(\chi_L,\hat{\theta})  W_L(z_L,z_G)d\chi_L\ .
\ea 
Here, $\hat{\theta}$ is the direction on the
sky. $\delta_m(\chi_L,\hat{\theta})$ is the matter overdensity (lens) at 
direction $\hat{\theta}$ and comoving angular distance $\chi_L\equiv \chi(z_L)
$  to redshift $z_L$.    $\chi_G\equiv \chi(z_G)$ is the comoving
angular diameter distance to the source. Both $\chi_L$ and $\chi_G$ are  in
unit of  $c/H_0$, where $H_0$ is the present day
Hubble constant. The lensing kernel 
\be
W_L(z_L,z_G)=\frac{3}{2}\Omega_m(1+z_L) 
\chi_L\left(1-\frac{\chi_L}{\chi_G}\right)\ .
\ee
when $z_L<z_G$ and is zero otherwise. $\Omega_m$
is the present day matter density in  unit of the critical density. The
lensing power spectrum is given by the following Limber equation,
\be
C^{GG}_{ij}(\ell)=\frac{\pi}{\ell}\int_0^{\infty}
\Delta^2_m(k=\frac{\ell}{\chi_L},z_L)W_i(z_L)W_j(z_L)\chi_L d\chi_L \ . 
\ee
Here, $W_i$ is the lensing kernel $W_L$ averaged over the $i$-th redshift bin,
defined by Eq. \ref{eqn:Wj}.

For two-point statistics, weak lensing surveys thus provide three sets of
correlations. Throughout this paper, we will work in the Fourier space
(multipole $\ell$ space) and
focus on  the  corresponding angular power spectra. The first one is the
angular cross correlation power 
spectrum between galaxy ellipticity  ($\kappa^s$) in the $i$-th redshift bin
and the one in the $j$-th redshift bin, $C^{(1)}_{ij}(\ell)$. 
\ba
\label{eqn:c1}
C^{(1)}_{ij}(\ell)=C^{GG}_{ij}(\ell)+C^{IG}_{ij}(\ell)+C^{IG}_{ji}(\ell)+C^{
II}_{ij}(\ell)\ .
\ea
Here, $C^{\alpha\beta}_{ij}$ is the angular cross correlation power spectrum
between quantity $\alpha$ in the $i$-th redshift bin and quantity $\beta$ in
the $j$-th redshift bin. $\alpha,\beta=G,I,g$, where the superscript (or
subscript in denoting the redshift and distance) $G$
denotes gravitational  lensing ($\kappa$), $I$  the galaxy intrinsic alignment
(non-random intrinsic ellipticity) and $g$ the galaxy number density
distribution  in the corresponding redshift bin ($\delta_\Sigma$).

The ellipticity-ellipticity pair relevant to the current paper has $i<j$.
Since $C^{IG}_{ji}\ll C^{IG}_{ij}$ as long as the catastrophic error is
reasonably small, we have 
\ba
\label{eqn:c11}
C^{(1)}_{ij}(\ell)\simeq C^{GG}_{ij}(\ell)+C^{IG}_{ij}(\ell)+2C^{II}_{ij}(\ell)\ \ {\rm when} \  i<j\ .
\ea
For bin size $\Delta z\ga 0.2$, as long as  the
catastrophic photo-z errors are reasonably small, $C^{II}_{ij}$ of
non-adjacent bins ($i<j-1$) is in general negligible with respect to the GI correlation
(e.g. \citealt{Schneider09}), because the II correlation only exists at small 
line-of-sight separation. However, for adjacent bins ($i=j-1$), the II
correlation can be non-negligible for $\Delta z\sim 0.2$
\citep{Schneider09}. The self-calibration technique proposed in the current
paper is not able to correct for the II correlation, for which other methods
\citep{Joachimi08,Joachimi09,Zhang10} may be applicable. It only applies to
correct for the GI correlation $C^{GI}$.  We express the GI
correlation as a fractional error to the lensing measurement,    
\be
\label{eqn:fij}
f_{ij}^I(\ell)\equiv \frac{C_{ij}^{IG}(\ell)}{C_{ij}^{GG}(\ell)}\ .
\ee 
$f_{ij}^I$ is the fractional GI contamination to the lensing
measurement. $f_{ij}^I$ ($i<j$) and $f_{ik}^I$ ($i<k$) are not independent, since both 
  describe the intrinsic alignment in the $i$-th redshift bin and thus 
\ba
\frac{f_{ij}^I(\ell)}{f_{ik}^I(\ell)} \simeq  
\left(\frac{W_{ij}}{W_{ik}}\right)\left(\frac{C_{ik}^{GG}(\ell)}{C_{ij}^{GG}
(\ell)}\right) \ .
\ea
However, for its clear meaning as a fractional error in the lensing
measurement, and for uncertainties in the intrinsic alignment modeling,
we adopt it, instead of $C^{IG}_{ij}$ itself,  to express the GI contamination throughout 
the paper.  The measured GI correlation is an
anti-correlation ($f_{ij}^{I}<0$), because the lensing induced shear is
tangential to the gradient of the  gravitational potential while the intrinsic
shear is parallel to the  gradient. Throughout the paper, we often neglect this
negative sign, since we will work under the limit $f_{ij}^I\ll 1$ and thus its sign
does not affect our error analysis.  Our self-calibration technique works in
principle for any value of $f_{ij}^I$.  The results presented in this paper
can be extended to other values of $f_{ij}^I$ straightforwardly. 

The second correlation is between the galaxy density ($\delta_{\Sigma}$) in the
$i$-th redshift bin and the galaxy ellipticity ($\kappa^s$) in the $j$-th redshift bin,
$C^{(2)}_{ij}$. Galaxy-galaxy lensing in general focuses on $C^{(2)}_{ij}$
($i<j$). \citet{Zhang09} showed that the measurement $C^{(2)}_{ij}$
($i>j$) contains valuable information on photo-z outliers.  The one
relevant for our GI self-calibration is 
\ba
\label{eqn:c2}
C^{(2)}_{ii}(\ell)=C_{ii}^{gG}(\ell)+C_{ii}^{gI}(\ell)\  \ .
\ea
The $C^{gI}_{ii}$ term contains here clearly contains valuable information
on  the intrinsic alignment.  The two terms on the right hand side has
different dependences on photo-z error. Larger photo-z errors tend to
increase $C^{gG}$ and decrease $C^{Ig}$. 

The third set of cross correlation is between galaxy density ($\delta_\Sigma$)
in the $i$-th  redshift bin and $j$-th redshift bin, $C^{(3)}_{ij}$. It has been shown that
$C^{(3)}_{ij}$ ($i\neq j$) is a sensitive probe of photo-z outliers
\citep{Schneider06,Zhang09}. Our GI 
self-calibration requires the measurement
\ba
\label{eqn:c3}
C^{(3)}_{ii}(\ell)=C_{ii}^{gg}(\ell) .
\ea

Our self-calibration aims to estimate and eliminate the GI contamination $C^{IG}_{ij}$ in
Eq. \ref{eqn:c11} from the measurements $C^{(2)}_{ii}$ (Eq. \ref{eqn:c2}) and
$C^{(3)}_{ii}$ (Eq. \ref{eqn:c3}), both are available in the same survey. This method is
thus dubbed the GI self-calibration.

For the self-calibration to work, $f_{ij}^I(\ell)$ must be sufficiently large
for the band power  $C^{Ig}_{ii}(\ell)$ at the corresponding $\ell$ bin to be
detected through the measurement $C^{(2)}_{ii}(\ell)$. We 
denote the threshold as  $f_{ij}^{\rm thresh}(\ell)$.  For brevity, we often
neglect the argument $\ell$ in $f_{ij}$. When $f_{ij}^I\geq
f_{ij}^{\rm thresh}$,  we can apply the self-calibration to reduce the GI
contamination. The residual GI contamination after the self-calibration is
expressed in as {\it residual fractional} error on the lensing measurement, in
which $\Delta f_{ij}$ denotes statistical error 
and $\delta f_{ij}$ denotes systematical error.  Thus the self-calibration performance is
quantified by $f_{ij}^{\rm thresh}$, $\Delta f_{ij}$ and $\delta
f_{ij}$. The smaller these quantities are, the better the performance. We will
numerically evaluate these quantities later.

\bfi{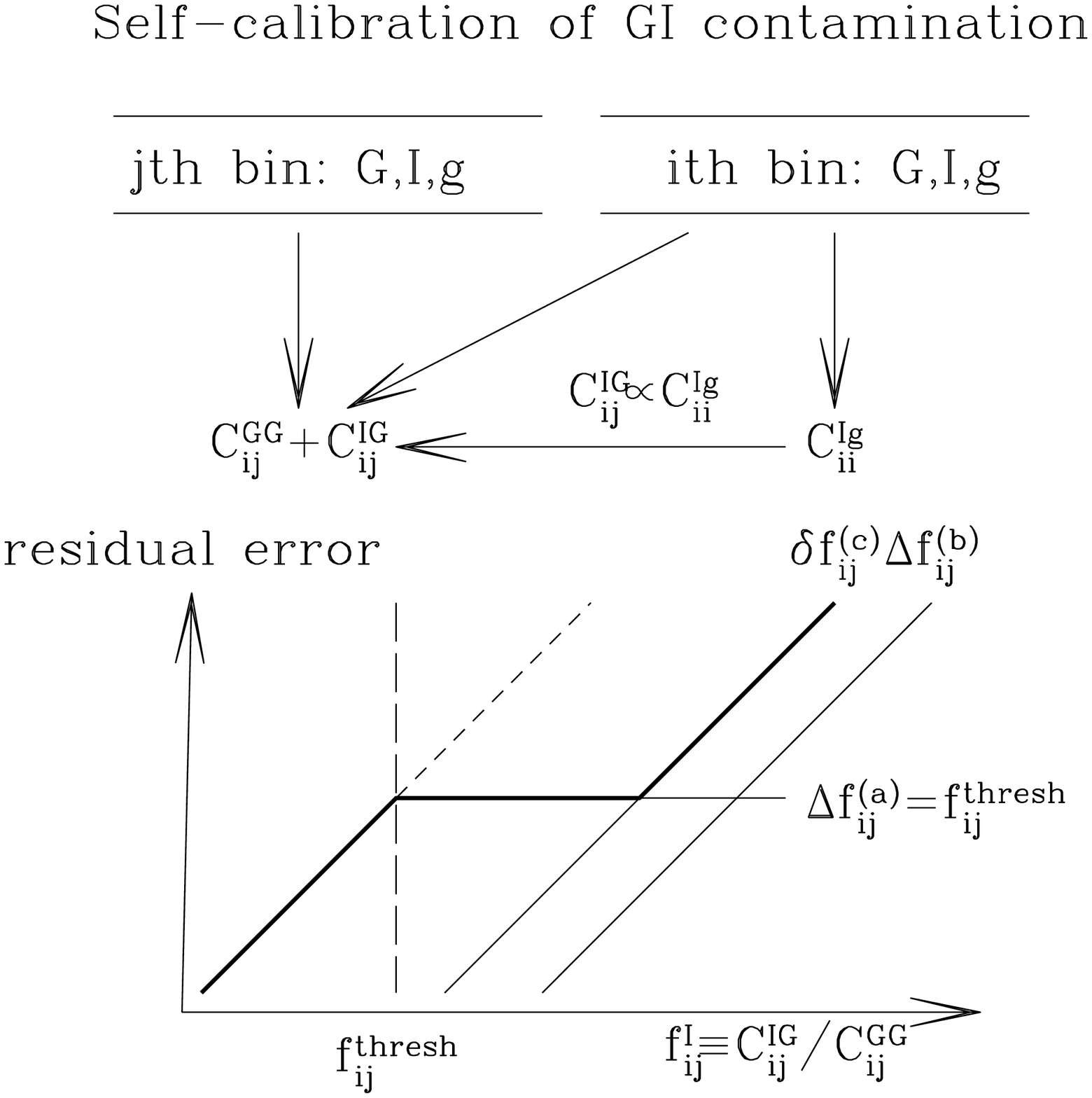}
\caption{A schematic description of the GI self-calibration technique. Here,
  $G$ denotes gravitational lensing, $I$  
  the non-random galaxy intrinsic alignment and $g$ the galaxy number density
  distribution  in the corresponding redshift bin. Here, galaxies  in the
  $j$-th redshift bin have higher photo-z than those   in the $i$-th redshift
  bin. The GI  contamination $C_{ij}^{IG}$ is expressed as fractional error in
  lensing power spectrum $C_{ij}^{GG}$, namely, $f^I_{ij}\equiv
  C_{ij}^{IG}/C_{ij}^{GG}$.  The self-calibration operates when $f_{ij}^I$ is
  bigger than certain threshold $f_{ij}^{\rm thresh}$.   Depending on the
  fiducial value of $f_{ij}^I$,  residual GI contamination from different
  sources dominates, which we highlight as bold lines in lower part of the
figure.  Refer to Eq. \ref{eqn:fij}, \ref{eqn:fija}, \ref{eqn:fijb} \&
\ref{eqn:fijc} for definitions of corresponding variables. 
\label{fig:1}}  
\efi

\subsection{The self-calibration}
The starting point of the GI self-calibration is a simple scaling relation
between $C^{IG}_{ij}$ and $C^{Ig}_{ii}$ that we find, 
\be
\label{eqn:relation}
C^{IG}_{ij}(\ell)\simeq \left[\frac{W_{ij}\Delta_i}
{b_i(\ell)} \right] C^{Ig}_{ii}(\ell)\ .
\ee
Here, $\chi_i\equiv \chi(\bar{z}_i)$, $b_i(\ell)$ is the galaxy bias in this
redshift bin at the corresponding angular scale $\ell$. $W_{ij}$ is the weighted lensing
kernel defined by 
\be
\label{eqn:wij}
W_{ij}\equiv \int_0^{\infty} dz_L \int_0^{\infty} dz_G
\left[W_L(z_L,z_G)\bar{n}_i(z_L)\bar{n}_j(z_G)\right]\ .
\ee
Notice that $\bar{n}_i$ is normalized such that
$\int_0^{\infty}\bar{n}_i(z)dz=1$. 
$\Delta_i$ is an effective width of the $i$-th redshift bin, defined by 
\be
\label{eqn:Deltai}
\Delta_i^{-1}\equiv \int_0^{\infty} \bar{n}_i^2(z) \frac{dz}{d\chi}dz\ .
\ee
 Refer to the appendix  \S \ref{sec:scaling} for the derivation. 

This scaling relation arises from the fact that, both GI and I-g cross
correlations are determined by the matter-intrinsic alignment in the $i$-th
redshift bin. The prefactors in Eq. \ref{eqn:relation} are simply the relative
weighting between the two. In this sense, this scaling relation is rather
generic. 

The basic procedure of the self-calibration, as illustrated in
Fig. \ref{fig:1},  is as follows. 
\bi
\item Extract $C^{Ig}_{ii}$ from the measurement
  $C^{(2)}_{ii}$ in the $i$-th photo-z bin. This exercise is non-trivial,
  since $C^{(2)}_{ii}$ actually measures  the sum of $C^{Ig}_{ii}$ and
  $C^{Gg}_{ii}$, and $C^{Gg}_{ii}$ is often non-negligible due to relatively
  large photo-z error. We find a method to
  simultaneously measure the two without resorting to spect-z information. The
  idea will be elaborated in  \S  \ref{subsec:Igmeasurement} and the
  measurement error in  $C^{Ig}_{ii}$ will 
  be calculated in  \S \ref{subsec:Ig}. 
\item Measure the galaxy bias $b_i(\ell)$ from the measurement $C^{(3)}_{ii}(\ell)$ (\S
  \ref{subsec:bias}).
\item Calculate $C^{IG}_{ij}$ from the above measurements and
  Eq. \ref{eqn:relation} and then subtract it from Eq. \ref{eqn:c1}.
\ei

\subsection{Measuring $C_{ii}^{Ig}$}
\label{subsec:Igmeasurement}
An obstacle in measuring $C_{ii}^{Ig}$ comes from the contamination
$C_{ii}^{Gg}$ (Eq. \ref{eqn:c2}). For spectroscopic sample,  this
contamination is straightforward to 
eliminate. We just throw away pairs where the redshift of the galaxy to
measure the shape is lower than the one to measure the number density 
\citep{Mandelbaum06,Hirata07,Okumura09}. This method is  robust, however
limited to  the spec-z sample.  

For photo-z sample, the above technique does not work, since the photo-z error
is large. For it, the true galaxy distribution, even for photo-z bin size $\Delta 
 z\rightarrow 0$,  has relatively large width $\geq \sim 0.1(1+z)$, no matter
 how narrow the photo-z bin is. In practice, the photo-z bin size is often
 $\ga 0.2$, which further increases the effective width. Thus, a galaxy in 
this photometric redshift bin has large chance to 
lens another galaxy in the same redshift bin, with a non-negligible lensing
weight. For this reason,  $C_{ii}^{Gg}$ may not be negligible comparing to
$C_{ii}^{Ig}$, even if the photo-z bin size is infinitesimal. We have
numerically compared $C_{ii}^{Gg}$ with $C_{ii}^{Ig}$ calculated based on the
intrinsic alignment model of  \citet{Schneider09} along their 
fiducial parameters, and found that the two terms can indeed be
comparable. For example, at a typical lensing angular scale $\ell=10^3$ and a
typical redshift $z=1$,  $C_{ii}^{Gg}/C_{ii}^{Ig}$ is $30\%$ for
vanishing bin size and increases with bin size. Only when the redshift is
sufficiently low or the redshift error and the bin size are both sufficiently
small, may the G-g correlation be safely neglected.

\bfi{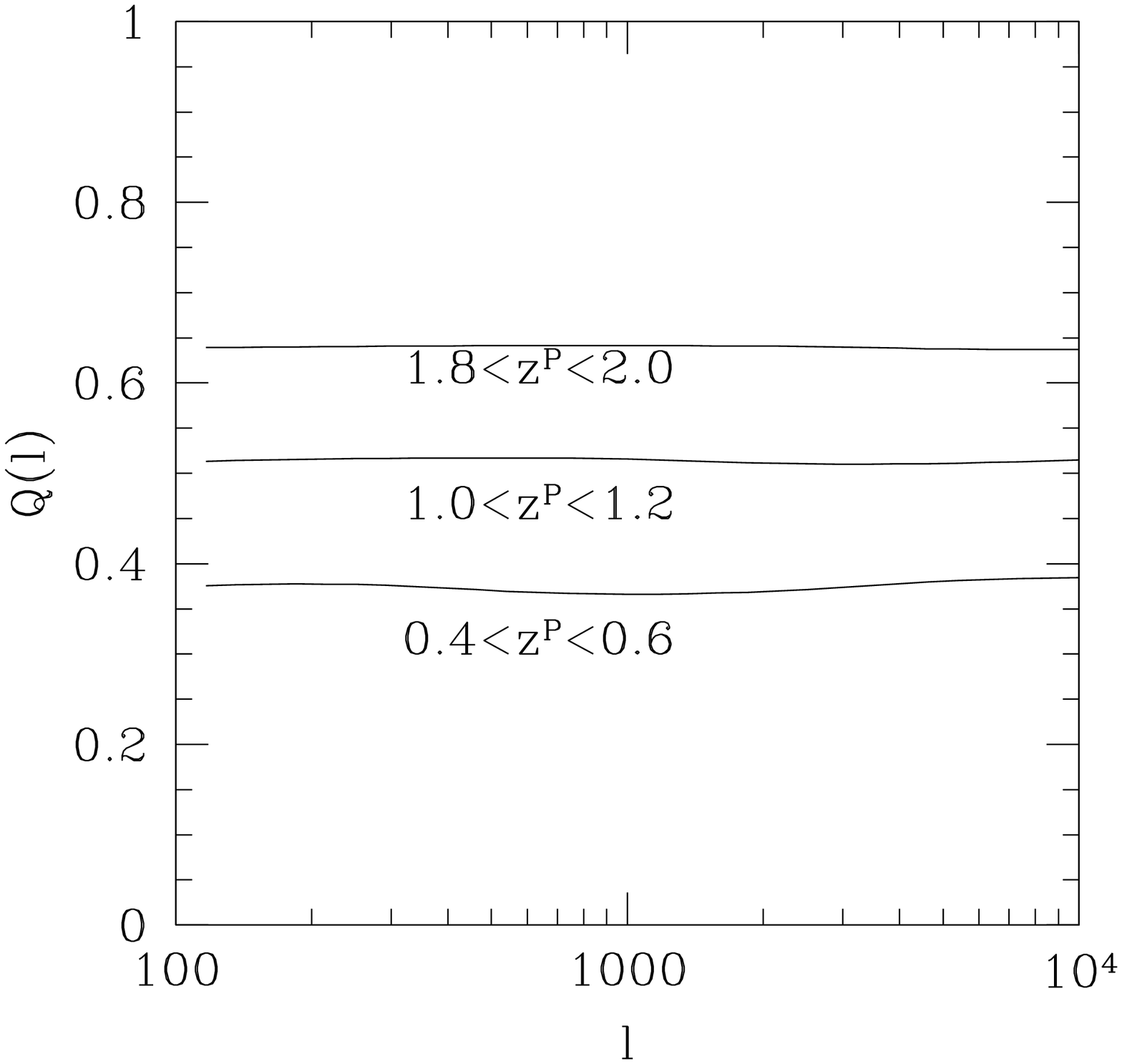}
\caption{$Q_i(\ell)\equiv C^{Gg}_{ii}|_S(\ell)/C^{Gg}_{ii}(\ell)$ describes
  the suppression of the galaxy-galaxy correlation in the same redshift bin
  after throwing away 
  those pairs with source redshift (photo-z) higher than lens redshift
  (photo-z). Only when  $Q$ deviates significantly from unity, can we extract
  the I-g correlation  from the shape-density correlation measurement in the
  same redshift bin. $Q$ 
  is nearly scale independent, since it is the ratio of two power spectra of
  similar shape.  Overall, $Q\sim 1/2$ .  It increases with redshift, as
  expected from larger photo-z rms error at higher redshift and hence larger
  effective redshift width.  This figure is a key result to demonstrate the
  feasibility of the self-calibration technique.  \label{fig:Q}}
\efi

Nevertheless, the photo-z measurement contains useful information
and  allows us to measure $C_{ii}^{Ig}$. Our method  to separate $C_{ii}^{Gg}$
from $C_{ii}^{Ig}$ relies on their
distinctive and predictable dependences on the relative position of galaxy
pair. Let's denote the redshift of one galaxy in the pair for the shape measurement as
$z_G^P$ and the other one for the number density measurement as $z_g^P$. The I-g
correlation does not depend on the ordering along the
  line-of-sight, as long as the physical
separation is fixed.  In another word, the I-g correlation for the pair with
$z_G^P>z_g^P$ is statistically identical to the pair with $z_G^P<z_g^P$, as
long as $|z_G^P-z_g^P|$ fixes.  On the other hand, the G-g correlation cares
about the ordering along the line-of-sight.  The G-g correlation for the pair with
$z_G^P>z_g^P$  is statistically larger than the pair with $z_G^P<z_g^P$, due
to the lensing geometry dependence. 

 We can then construct two observables on the ellipticity-density
 correlation. (1) One is $C_{ii}^{(2)}(\ell)$, 
weighting all pairs equally. (2) The other is  $C_{ii}^{(2)}|_S(\ell)$, in
which we only count the cross correlation between those pairs with
$z_G^P<z_g^P$. The subscript ``$S$'' 
denotes the corresponding quantities under this weighting. From the argument
above,  we have $C_{ii}^{Ig}(\ell)=C_{ii}^{Ig}|_S(\ell)$. On the
other hand,  $C_{ii}^{Gg}|_S(\ell)<C_{ii}^{Gg}(\ell)$, since those
$z_G^P>z_g^P$ pairs that we disregard contribute
more to the lensing-galaxy correlation.  We denote the suppression by 
\be
Q(\ell)\equiv \frac{C_{ii}^{Gg}(\ell)|_S}{C_{ii}^{Gg}(\ell)}\ .
\ee
$Q$ is sensitive to photo-z errors. In general catastrophic errors drive $Q$ 
towards $1$. $Q=1$ if the photo-z is completely wrong
and has no correlation with the true  redshift  and $Q=0$ if the photo-z is
$100\%$ accurate. Usually $0<Q<1$.  
$Q$ can be calculated from the galaxy redshift  distribution (the appendix \S
\ref{sec:Q}) and is thus in principle an observable too. From the  two observables
\ba
\label{eqn:pair}
C_{ii}^{(2)}(\ell)&=&C_{ii}^{Ig}(\ell)+C_{ii}^{Gg}(\ell)\ \ ,\nonumber \\
C_{ii}^{(2)}|_S(\ell)&=&C_{ii}^{Ig}(\ell)+C_{ii}^{Gg}|_S(\ell)  \ \ ,
\ea
We obtain the solution to $C_{ii}^{Ig}$, 
\be
\label{eqn:estimator}
\hat{C}_{ii}^{Ig}(\ell)=\frac{C^{(2)}_{ii}|_S(\ell)-Q(\ell)C^{(2)}_{ii}
(\ell)}{1-Q(\ell)}\ .
\ee
For it to be non-singular,
$Q$ must deviate from unity ($Q<1$). We numerically evaluate this quantity and find
that, for the survey specifications presented in this paper, $Q\sim 1/2$
(Fig. \ref{fig:Q}). The significant deviation of $Q$ from unity is not a
coincidence. It is in fact a manifestation that the lensing efficiency changes
significantly across the redshift interval $\sigma_P$. We thus expect the
$C_{ii}^{Ig}$ estimator (Eq. \ref{eqn:estimator}) to be  applicable in general.

\section{Error estimation}
\label{sec:error}
Unless explicitly specified,  we will target at LSST throughout this 
paper to estimate the performance of the self-calibration technique proposed
above.  LSST plans to cover half the sky ($f_{\rm sky}=0.5$) and reach the
survey depth of  $\sim 40$ galaxies per arcmin$^{2}$. We  
adopt the galaxy number density distribution as  $n(z)\propto
z^2\exp(-z/0.5)$,  the rms shape error $\gamma_{\rm rms}=0.18+0.042z$ and
photo-z scatter $\sigma_P=0.05(1+z)$.  We split galaxies into photometric
redshift bins  with $\Delta z_i=0.2$ centered   at  $\bar{z}_i=0.2(i+1)$
($i=1,2,\cdots$).\footnote{The choice of redshift bin is somewhat arbitrary. For
  example, we can include a lower redshift bin with $\bar{z}_i=0.2$
  ($0.1<z^P<0.3$).  The self-calibration certainly works for this bin, since
  the I-g cross correlation in this bin is easier to extract due to weaker
  lensing signal and hence weaker G-g in this bin. The major reason that we do not
  include this bin is that the lensing signal in this bin is weak. Weak
  lensing signal may cause confusion on the performance of the
  self-calibration, due to our choice to express the GI correlation before and
  after the self-calibration as ratios with respect to the lensing signal. For
  example, Fig. \ref{fig:fij} shows that $\Delta f_{ij}$ increases toward low
  redshifts. But it is an manifestation of weak lensing signal instead of
  poor performance of the self-calibration.  } 

\subsection{Measurement error in $C_{ii}^{Ig}$}
\label{subsec:Ig}

\bfi{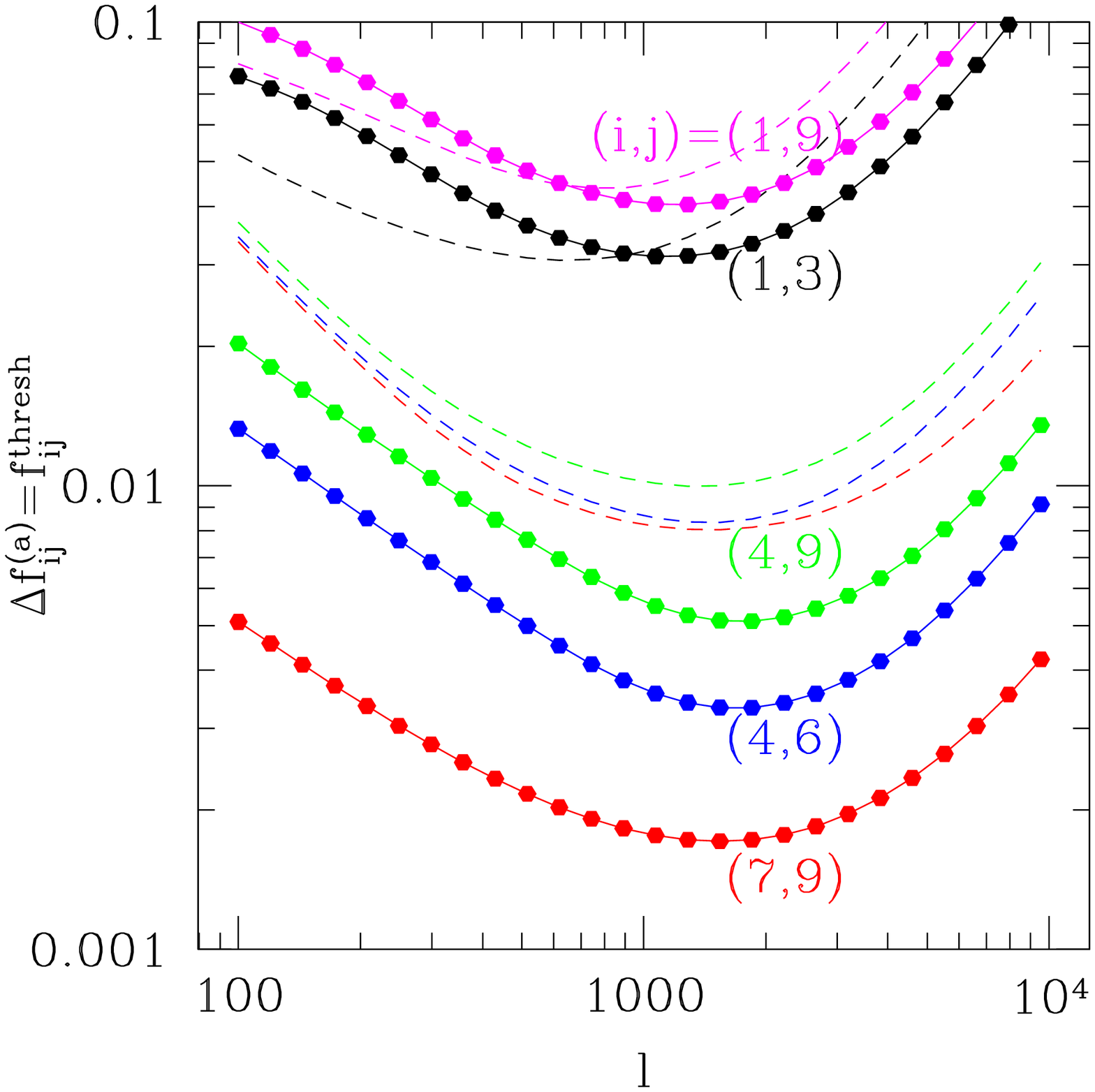}
\caption{The threshold of the applicability of the self-calibration technique
  $f_{ij}^{\rm thresh}$ and the residual statistical uncertainty $\Delta
  f_{ij}^{(a)}$ (solid lines with data points). Refer to Eq. \ref{eqn:fija}
  for the definition of $\Delta f_{ij}^{(a)}$. The self-calibration applies 
  for sufficiently large GI contamination ($f_{ij}^I> f_{ij}^{\rm thresh}$)
  and reduces the fractional error from $f_{ij}^I$ to $\Delta
  f_{ij}^{(a)}$. Here, the superscript $(a)$ denotes the error from the
$C^{Ig}_{ii}$
measurement.  Notice that $f_{ij}^{\rm thresh}=\Delta f_{ij}^{(a)}$ and both
are insensitive to the fiducial $f_{ij}^I$.  The cosmic variance in the
lensing field and the random shape fluctuation sets the minimum {\it
  fractional} statistical error $e^{\rm min}_{ij}$ in $C^{GG}_{ij}$
measurement, which we plot as dash lines. The  numerical estimation shown  is
for redshift bins $\bar{z}_i=0.2(i+1)$ ($i=1,2,\cdots$) and multipole bin size
$\Delta \ell=0.2 \ell$ in LSST. In general $\Delta f_{ij}^{(a)}<e^{\rm
  min}_{ij}$ and thus the residual error after the self-calibration is
negligible.  Since both  $\Delta f_{ij}^{(a)}$ and $e_{ij}^{\rm min}$ scale 
in similar ways, this conclusion holds in general for other lensing
surveys.    \label{fig:fij}}    
\efi

Both $C_{ii}^{(2)}$ and  $C_{ii}^{(2)}|_S$ have cosmic variance and shot noise
errors, which propagate into $C_{ii}^{Ig}$ extracted from the estimator
Eq. \ref{eqn:estimator}.  The error estimation on $C_{ii}^{Ig}$  is
non-trivial since errors in $C_{ii}^{(2)}$ and  $C_{ii}^{(2)}|_S$ are neither
completely uncorrelated, nor completely correlated. The derivation is lengthy,
so we leave it  in the appendix \S \ref{sec:appendixC}.   The final result of
the rms error $\Delta C_{ii}^{Ig}$ in a bin with width $\Delta \ell$ is 
\ba
\label{eqn:cigerror}
(\Delta
C_{ii}^{Ig})^2&=&\frac{1}{2\ell\Delta \ell f_{\rm
    sky}}\left(C^{gg}_{ii}C^{GG}_{ii}+\left[1+\frac{1}{3(1-Q)^2}\right]\right.
\no
\\
&&\times
\left[C^{gg}_{ii}C^{GG,N}_{ii}+C^{gg,N}_{ii}(C^{GG}_{ii}+C^{II}_{ii})\right]\no
\\
&&\left.+C^{gg,N}_{ii}C^{GG,N}_{ii}\left[1+\frac{1}{(1-Q)^2}\right] \right)
\ .
\ea
Here, the superscript ``N'' denotes measurement noise such as shot noise in
galaxy distribution and random shape noise. $C^{gg,N}_{ii}=4\pi f_{\rm
  sky}/N_i$ and $C^{GG,N}_{ii}=4\pi f_{\rm
  sky}\gamma_{\rm rms}^2/N_i$, where $N_i$ is the total number of galaxies in
the $i$-th redshift bin. 

From Eq. \ref{eqn:cigerror}, $\Delta C^{Ig}_{ii}$ is insensitive to the
intrinsic alignment, as long as the II correlation is sub-dominant with respect to the GG
correlation. We will work at this limit.  
If the intrinsic alignment is too small, the measurement error
$\Delta C_{ii}^{Ig}$ will be larger than $C^{Ig}_{ii}$.  $\Delta
C_{ii}^{Ig}=C_{ii}^{Ig}$  hence sets a threshold $f_{ij}^I$.
Combining  Eq. \ref{eqn:relation} and the definition
$f_{ij}^I\equiv C^{IG}_{ij}/C^{GG}_{ij}$ (Eq. \ref{eqn:fij}), we obtain this threshold as
\be
\label{eqn:threshold}
 f_{ij}^{\rm thresh}=\left(\frac{\Delta   
C_{ii}^{Ig}}{C_{ij}^{GG}}\right)\left(\frac{W_{ij}\Delta_i}{b_i(\ell)}\right)
 \ .
\ee

$f_{ij}^{\rm thresh}$  has two important meanings. (1) It describes the
minimum intrinsic alignment ($f_{ij}^I=f_{ij}^{\rm thresh}$) that can be detected through our
  method with S/N=1. Thus it also defines the lower limit beyond which  our
self-calibration technique is no longer applicable. 
(2) It describes the self-calibration accuracy resulting from $C^{Ig}_{ii}$
  measurement error. The measurement error $\Delta
  C_{ii}^{Ig}$ propagates into an error in $C_{ij}^{IG}$ determination through
  Eq. \ref{eqn:relation} and hence leaves a residual statistical error in the
  GG measurement. We denote this error  as $\Delta f_{ij}^{(a)}$. Since $\Delta
  f_{ij}^{(a)}/f_{ij}^I= \Delta C_{ii}^{Ig}/ C_{ii}^{Ig}$, combining
  Eq. \ref{eqn:relation}, we have 
\be
\label{eqn:errora}
\Delta f_{ij}^{(a)}= \left(\frac{\Delta
C_{ii}^{Ig}}{C_{ij}^{GG}}\right)\left(\frac{W_{ij}\Delta_i}{b_i(\ell)}\right) \
.
\ee
Also, we find an important relation between the two,
\be
\label{eqn:fija}
\Delta f_{ij}^{(a)}= f_{ij}^{\rm thresh}\ . 
\ee
This relation can be understood in this way. When the intrinsic alignment
$f^{I}_{ij}>f^{\rm thresh}_{ij}$, it can be inferred through
measurement in $C^{Ig}_{ii}$ and hence be corrected. Since measurement  $C^{Ig}_{ii}$
has statistical error $\Delta C^{Ig}_{ii}$ (Eq. \ref{eqn:cigerror}), this
correction is imperfect. The residual error in $f^{I}_{ij}$ is set by the same
$\Delta C^{Ig}_{ii}$ determining $f^{\rm thresh}$, so we have the relation
Eq. \ref{eqn:fija}.  If only our method is applied,  we are only able to
detect the intrinsic alignment with an amplitude $f^I_{ij}>f^{\rm thresh}$ and correct it to a
level of $\Delta f^{(a)}_{ij}=f^{\rm thresh}$. 

Once the intrinsic alignment is sufficiently large to be detected
($f_{ij}^I>f_{ij}^{\rm thresh}$), our self-calibration technique can detect and
thus eliminate the  GI contamination. It  renders a systematical error in
lensing measurement with amplitude $f_{ij}^I$ into a statistical error with
rms $\Delta f_{ij}^{(a)}=f_{ij}^{\rm thresh}<f_{ij}^I$.  We notice that
$\Delta f_{ij}^{(a)}$ is insensitive to $f_{ij}^I$. 

Numerical results on $\Delta f_{ij}^{(a)}=f_{ij}^{\rm thresh}$ are evaluated
through Eq. \ref{eqn:errora} along with Eq. \ref{eqn:cigerror} and are shown in
Fig. \ref{fig:fij}. For comparison, we also plot {\it the minimum G-G error} in
the 
$C_{ij}^{GG}$ measurement. It is the rms fluctuation  induced by the cosmic
variance in the G-G correlation and shot  noise due to random galaxy ellipticities,  in
idealized case of no other error sources such as the  intrinsic
alignment.  This minimum G-G error sets the ultimate lower limit of the
fractional  measurement error on $C^{GG}_{ij}$ ($i\neq j$), 
\be
\left(e_{ij}^{\rm min}\right)^2=\frac{C^{GG
,2}_{ij}+(C^{GG}_{ii}+C^{GG,N}_{ii})(C^{GG}_{jj}+C^{GG,N}_{jj})}{2\ell \Delta \ell f_{\rm
sky} C^{GG
,2}_{ij}}\ . 
\ee
When $\Delta f_{ij}^{(a)}$ is smaller than $e_{ij}^{\rm min}$, the
residual error after the self-calibration will then be negligible, meaning a 
self-calibration with little cosmological information loss. We find that this is
indeed the case in general. 
Fig. \ref{fig:fij} is the forecast for LSST.\footnote{We do notice that in
  some cases especially when one of the photo bin is at low redshift,  $\Delta
  f_{ij}^{(a)}>e_{ij}^{\rm min}$ (Fig. \ref{fig:fij}), leading to
  non-negligible loss in cosmological information for relevant redshift bins.} Since both $\Delta f_{ij}^{(a)}$
and $e_{ij}^{\rm min}$ scale in a similar way with respect
to survey specifications such as the sky coverage and galaxy number density,
this conclusion also holds 
for other lensing surveys such as CFHTLS, DES, Pan-Starrs, Euclid and JDEM. 

How can the GI
contamination be corrected to an accuracy even below the statistical limit of
GG measurement? Equivalently, why is $f_{ij}^I$ as small as the one shown in
Fig. \ref{fig:fij} detectable? This surprising result requires some
explanation.  The reason is that $C_{ii}^{Ig}$ is amplified with respect to
$C_{ij}^{IG}$ by  a large factor $1/W_{ij}\Delta_i\sim O(10^2)$
  (Eq. \ref{eqn:relation}).  Thus a small GI contamination ($f_{ij}^I\ll 1$) can
still  cause a detectable $C_{ii}^{Ig}$. This explains  the small $f_{ij}^{\rm
  thresh}=\Delta f_{ij}^{(a)}$. 

\subsection{Measuring the galaxy bias}
\label{subsec:bias}
The second uncertainty in the self-calibration comes from the
measurement of the galaxy bias, $\Delta b_i(\ell)$. From
Eq. \ref{eqn:relation}, this uncertainty induces a
residual statistical error
\be
\label{eqn:fijb}
\Delta  f_{ij}^{(b)}=f_{ij}^I (\Delta b_i(\ell)/b_i(\ell))\ .
\ee

There are several possible ways to obtain $b_{\ell}$ such as combining
galaxy-galaxy
lensing and galaxy clustering measurements (e.g. \citealt{Sheldon04}), 
combining 2-point and 3-point
galaxy clustering (e.g. \citep{Guo09}), fitting against the halo occupation
\citep{Zheng05} and 
conditional luminosity function \citep{Yang03}, or analyzing the
counts-in-cells measurements \citep{Szapudi04}. Alternatively,  $b_i(\ell)$ can
be
inferred from the galaxy  density-galaxy density correlation measurement
alone, $
C^{(3)}_{ii}(\ell)=C^{gg}_{ii}(\ell)\simeq
b_{\ell}^2 C^{mm}_{ii}(\ell)$. 
Here, $C^{mm}_{ii}$ is the matter angular power spectrum, with the same
weighting as galaxies.  Both uncertainties in the theoretical prediction of
$C^{mm}_{ii}$  and measurement error in $C^{(3)}_{ii}$ affects the $b_i(\ell)$
measurement.  Given a cosmology, one can evolve the matter power spectrum
tightly constrained at the recombination epoch by CMB experiment to low
redshift and thus predict $C^{mm}_{ii}$. As long as the associated uncertainty
is smaller than $10\%$, the induced error will be sub-dominant to the
systematical error discussed later in \S \ref{subsec:systematic}.   This is an
important issue for further investigation.  

On the other hand,  the statistical 
error in $b_i(\ell)$ induced by the galaxy clustering measurement error is  
\be
\label{eqn:bg}
\frac{\Delta b_i(\ell)}{b_i(\ell)}\sim
\frac{1}{2}\sqrt{\frac{1}{\ell\Delta
    \ell f_{\rm sky}}} \times
\left(1+\frac{C_{ii}^{gg,N}(\ell)}{C_{ii}^{gg}(\ell)}\right)\ .
\ee
This rough estimation suffices for the purpose of this paper, for which 
the reason  will be come clear latter soon. This error 
  is negligible for a number of reasons. (1) It is in general much smaller than
the
  systematical error in the scaling relation, $\delta f_{ij}^{(c)}$. As will
  be shown in \S \ref{subsec:systematic}, $\delta f_{ij}^{(c)}\sim
  0.1f_{ij}^I$. On other other hand,  $b_i(\ell)$ can in general be measured
with
  better than $10\%$ accuracy. For example, for LSST at $\ell=100$ with
  $\Delta \ell =0.2\ell$, $\Delta b_i(\ell)/b_i(\ell)\simeq 1.6\%$.  $b_i(\ell)$
can be
measured
  with higher accuracy toward smaller scales, until shot noise dominates. Thus
  $\Delta f_{ij}^{(b)}\ll \delta f_{ij}^{(c)}$ at relevant scales. (2) It is
  smaller than the minimum statistical error in  $C_{ij}^{GG}$ 
  measurement (Fig. \ref{fig:fij}).  First of all, galaxy clustering measurement
  is in general more accurate than lensing measurement. Second, the impact of
  uncertainty in $b_i(\ell)$ on the self-calibration is modulated by a factor
  $f_{ij}^I$ ($\Delta f_{ij}^{(b)}=f_{ij}^I (\Delta b_i(\ell)/b_i(\ell)$)).
Unless $f_{ij}^I>1$, $\Delta f_{ij}^{(b)}$ is suppresed. Thus we expect that the
  $b_i(\ell)$ induced error $\Delta f_{ij}^{(b)}$ is negligble even to
the minimum statistical error in  $C_{ij}^{GG}$  measurement.  From the above
  general argument, this conclusion should hold for most, if not all, lensing
  surveys. We show one example of LSST. Even for a rather large 
  $f_{ij}^I=1$, uncertainty in $b_i(\ell)$ only causes a statistical error of 
  $\Delta f_{ij}^{(b)}=1.6\%$ at $\ell=100$,  negligible comparing to
statistical uncertainties in $C_{ij}^{GG}$ measurement
(Fig. \ref{fig:fij}). The above conclusions are safe even if Eq. \ref{eqn:bg}
underestimates the error in $b_i(\ell)$ by a factor of a few.  This is the
reason that we do not seek a more robust estimation on 
  the measurement error in $b_i(\ell)$. 

From the above argument, the errors $\Delta f_{ij}^{(a)}$ and $\Delta
f_{ij}^{(b)}$ arise from different sources and thus are independent to each
other. The combined error is then $\sqrt{(\Delta f_{ij}^{(a)})^2+(\Delta
  f_{ij}^{(b)})^2}$.  Since the galaxy bias induced error is likely
sub-dominant to either the 
error source $(a)$ or to the one will be discussed in \S
\ref{subsec:systematic},  we will neglect it for the rest of the paper.

\subsection{The accuracy of the $C^{IG}_{ij}$-$C^{Ig}_{ii}$ relation}
\label{subsec:systematic}
A key ingredient in the self-calibration is Eq. \ref{eqn:relation}, which links
the observable $C_{ii}^{Ig}$ to the GI contamination. However,
Eq. \ref{eqn:relation} is not exact and we quantify its accuracy by 
\be
\epsilon_{ij}(\ell)\equiv  \frac{b_i(\ell) C^{IG}_{ij}(\ell)}{W_{ij}\Delta_i
C^{Ig}_{ii}(\ell)}-1\ .
\ee
$\epsilon_{ij}$ also quantifies the induced residual systematic error, 
\be
\label{eqn:fijc}
\delta f_{ij}^{(c)}=\epsilon_{ij}f_{ij}^I\ .
\ee
We present a rough estimation by adopting a toy model
\be
\label{eqn:toy}
\Delta^2_{mI}(k,z)\propto \Delta^2_{gI} (k,z)\propto \Delta^2_m (k,z)
(1+z)^{\beta}
\ee with $\beta=-1,0,1$. Here, $\Delta^2_{mI}$, $\Delta^2_{gI}$ and
$\Delta^2_m$ are the 3D 
matter-intrinsic alignment, galaxy-intrinsic alignment  cross correlation
power spectrum (variance) and matter power spectrum, respectively. The
accuracy of Eq.  \ref{eqn:relation} is not only affected by the scale
dependence of corresponding 3D power spectra, but also their redshift
evolution (the appendix  \ref{sec:scaling}). Theoretical models of the
intrinsic alignment (e.g. \citealt{Schneider09}) show that the redshift
evloution may not follow the evolution in the density field. For this reason,
we add an extra redshift dependence $(1+z)^\beta$ in Eq. \ref{eqn:toy}. This
recipe is completely arbitrary, but it helps to
demonstrate the robustness of  Eq. \ref{eqn:relation}. 

Numerical results are shown in
Fig. \ref{fig:epsilon}. We see that for most $ij$ pairs, $|\epsilon_{ij}|$ is
less than $10\%$,  meaning that we are able to suppress
the GI contamination by a factor of $10$ or larger, if other errors are
negligible. 

\bfi{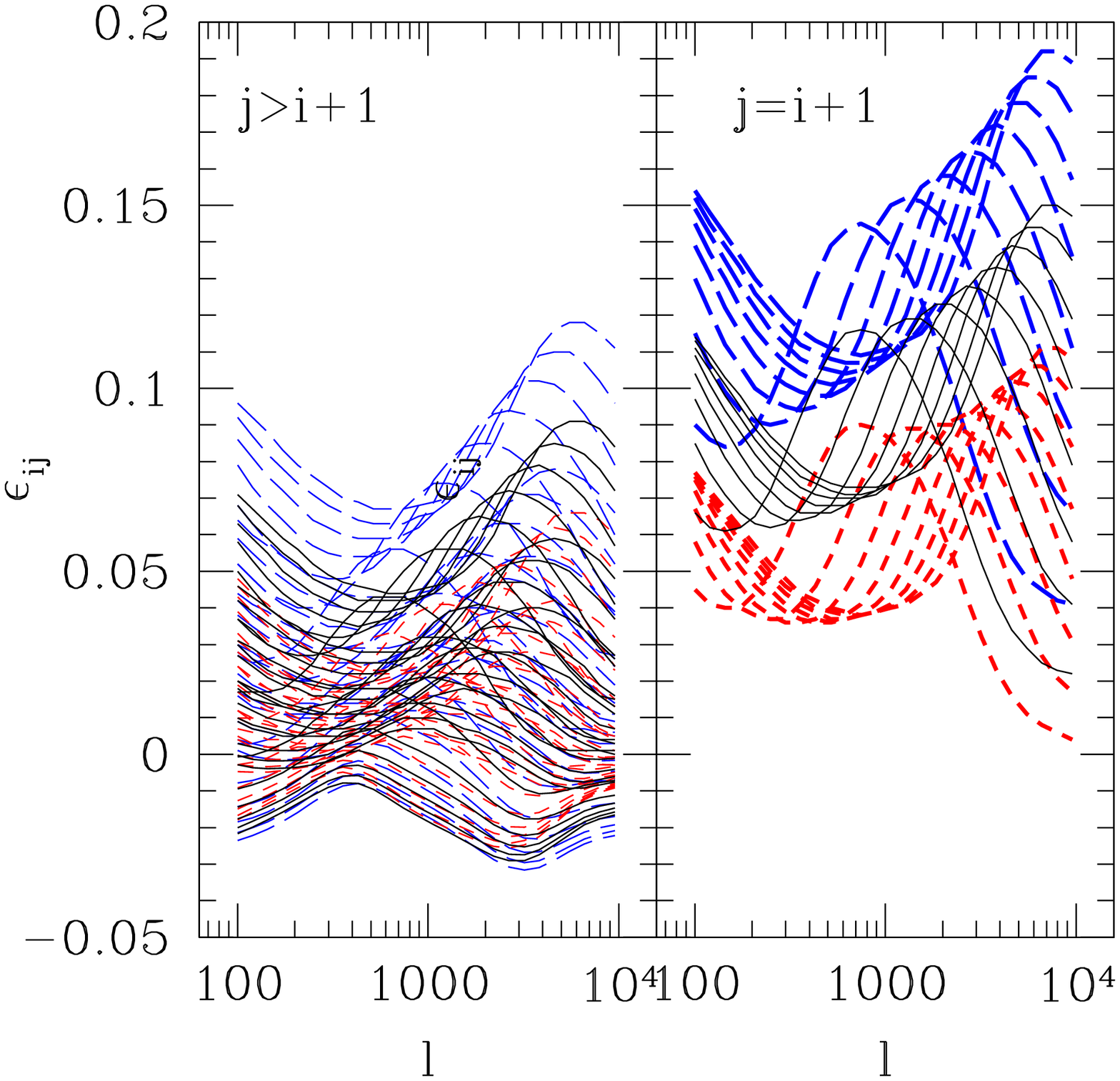}
\caption{$\epsilon_{ij}$ quantifies  the accuracy of Eq. \ref{eqn:relation},
  which links the observable $C_{ii}^{Ig}$ to the GI contamination
  $C_{ij}^{IG}$. It  thus quantifies a dominant systematical error of the
  self-calibration technique, $\delta f_{ij}^{(c)}=\epsilon_{ij} f_{ij}^I$.
Solid, dotted and dash lines
  represent three toy models of the intrinsic ellipticity evolution. Usually,
  Eq. \ref{eqn:relation} is accurate within $10\%$. However, for those
  adjacent bins with $i,j=i+1$, due to stronger redshift dependence  of the
  lensing kernel, Eq. \ref{eqn:relation} is least accurate and $\epsilon_{ij}$
  is largest (right panel).  \label{fig:epsilon}}  
\efi

We notice that the largest inaccuracy of Eq. \ref{eqn:relation}
occurs for those adjacent bins ($i,j=i+1$), which is $10\%$-$20\%$. This is
caused by the lensing geometry dependence. Eq. \ref{eqn:relation} would be quite accurate if the
integrand in Eq. \ref{eqn:IG} \& \ref{eqn:Ig} varies slowly with
redshift. However, since the $i$-th bin and the $j=i+1$-th bin are close in
redshift, the lensing kernel $W_j(z)$ varies quickly over the $i$th redshift
bin. This fast redshift evolution degrades the accuracy of the scaling
relation and thus causes a larger $|\epsilon_{ij}|$. On the
other hand,  $W_j(z)$ changes more slowly over other bins with $i\neq j-1$,
since now
the source-lens separation is larger. For this
reason, Eq. \ref{eqn:relation} is more accurate for these bins. The
self-calibration technique does not work excellently for adjacent bins, but a
factor  of $5$ reduction in the GI contamination is still achievable. 

The scaling relation accuracy is senstive to the photo-z
accuracy, $\epsilon_{ij} \propto  \sigma^2\simeq \sigma_P^2+(\Delta
z)^2/12$. Here, $\sigma$ is the rms redshift dispersion in the corresponding
redshift bin. One can obtain the above relation by perturbing 
Eq. \ref{eqn:IG} \&  \ref{eqn:Ig} around the median redshift and keeping up to
the second order.   Thus, if photo-z accuracy can be significantly improved,
the accuracy of Eq. \ref{eqn:relation} can be significantly improved too. For
example, if   $\sigma_p=0.03(1+z)$ instead of the fiducial value $0.05(1+z)$,
the scaling relation accuracy can be improved by a factor of $\sim
2$-$3$. This would allow us to suppress the GI contamination by a factor of
$10$-$20$ or even higher. The shape of the photo-z PDF also matters. Depending
on which direction it is skewed, the scaling relation accuracy may be improved
or degraded.  In \S \ref{subsec:photoz}, we will discuss the impact of
catastrophic error, which presents as significant deviation from the adopted
Gaussian PDF. 

The accuracy of the scaling relation may also be improved by better modeling. 
$\epsilon_{ij}$ has (much) larger chance  to be positive
(Fig. \ref{fig:epsilon}). This behavior is likely general, not limited to the
intrinsic  ellipticity toy models we investigate.  In deriving the scaling relation, both
$C^{Ig}_{ii}$ and $C^{IG}_{ij}$ are evaluated at the middle redshift
$\bar{z}_i$. The weighting function in $C^{Ig}_{ii}$ roughly peaks at
$\bar{z}_i$ while the one in $C^{IG}_{ii}$ peaks at lower redshift, due to the
monotonous decreasing of $W_j(z)$ with redshift (until it vanishes). It is this imbalance causing the general
behavior $\epsilon_{ij}>0$. It also shed ligh on improving 
$C^{IG}_{ij}$-$C^{Ig}_{ii}$ relation  and  reduceing the associated
systematical error: an interesting project for future investigation. 

\subsection{General behavior of the residual errors}
Depending on the nature of the GI correlation, either the error in $C^{Ig}_{ii}$
measurement or the error in the scaling relation can dominate the error budget
of the self-calibration, while the one from the galaxy bias is likely
sub-dominant.  Fig. \ref{fig:1} demonstrates the relative behavior 
of the three error sources in the self-calibration. The bold lines highlight
the dominant error, as a function of the intrinsic alignment amplitude
$f_{ij}^I$.  There are several regimes. 
\bi
\item $f_{ij}^I\leq f_{ij}^{\rm thresh}$. The intrinsic alignment is too weak
  to be detected in the galaxy-lensing correlation. For this reason, the
  self-calibration technique is not applicable.  However, this usually also
  mean the intrinsic alignment is negligible in lensing-lensing measurement,
  comparing to the associated minimum statistical error (Fig. \ref{fig:fij}).
  Thus there is no need to correct for the GI contamination in this
  case. However, there are important exceptions to the above conclusion. For
  example, from Fig. \ref{fig:fij}, when one photo-z bin is at sufficiently
  low redshift, the GI contamination is undetectable by our method, but the
  systematical error it induces is non-negligible. Furthermore, our method is
  insensitive to the intrinsic alignment which is weakly correlated to the
  density field. Such intrinsic alignment can cause large contamination to
  the lensing power spectrum $C^{GG}_{ii}$, but leaves no detectable feature
  in the ellipticity-density measurement $C^{(2)}_{ii}$. In these cases, other
  methods (e.g. \citealt{Joachimi08,Joachimi09,Zhang10})
  shall be applied to correct for the intrinsic alignment.

\item  $f_{ij}^I> f_{ij}^{\rm thresh}$. The self-calibration begins to work. 

(1) $f_{ij}^I<\Delta f_{ij}^{\rm thresh}/\epsilon_{ij}$. The statistical error
  induced by  I-g measurement uncertainty dominates. However, this residual
  error, $\Delta f_{ij}^{(a)}=\Delta f_{ij}^{\rm thresh}$, is usually
negligible,  comparing to the 
  minimum statistical error $e^{\rm min}_{ij}$ in shear measurement ($\Delta
  f_{ij}^{(a)}<e^{\rm min}_{ij}$, Fig. \ref{fig:fij}).  In
  this domain, the self-calibration technique is promising to work down to the
  statistical limit of lensing surveys. 

(2) $f_{ij}^I>\Delta f_{ij}^{\rm thresh}/\epsilon_{ij}$. The systematical error
  arising from the imperfect scaling relation domiantes. The fractional residual
error in 
  lensing-lensing measurement is $\delta f_{ij}^{(c)}=\epsilon_{ij}f_{ij}^I\sim
  0.1 f_{ij}^I$. This error will be still sub-dominant to the lensing statistical
  fluctuation, if $f_{ij}^I<e^{\rm min}_{ij}/\epsilon_{ij}$. If not the case,
  the self-calibration can work to suppress the GI contamination
  by a factor of $~10$. Other complementary techniques such as the nulling
  technique proposed by \citet{Joachimi08,Joachimi09} shall be applied to
  further reduce the residual GI contamination down to its statistical limit. 
\ei

\section{Other sources of error}
\label{sec:moreerrors}
There are other sources of error, beyond the three ones discussed above. We
discuss qualitatively on magnification bias, catastropic photo-z error,
stochastic galaxy bias and cosmological uncertainties. Based on  simplified
estimations,  we conclude that none
of them  can completely invalidate the self-calibration technique. Quantitative
and comprehensive evaluation of all these errors, 
including the ones in previous section,  will be postpone elsewhere. 
\subsection{Magnification bias}
Gravitational lensing not only distorts the shape of galaxies, but also alter
their spatial distribution and induces  magnification bias, or cosmic
magnification (e.g. \citet{Scranton05} and references therein). It changes the
observed galaxy overdensity to  
$\delta_g^L=\delta_g+g(F)\kappa$. The function $g(F)=2(-d\ln N(>F)/d\ln F-1)$
is determined by the logrithamic slope of the (unlensed) galaxy luminosity
function
$N(>F)$ and is in principle measurable. 

The magnification bias affects both $C^{(1)}_{ij}$, through a subtle  source-lens 
coupling \citep{Hamana01},  and $C^{(3)}_{ij}$. However, these impacts are
negligible, in the context  of this paper. The magnification bias has a
relatively larger effect on $C^{(2)}_{ii}$ and modifies Eq. \ref{eqn:c2}  to
\ba
C^{(2)}_{ii}&=&C_{ii}^{gG}+C_{ii}^{gI}+g_i(C_{ii}^{GG}+C_{ii}^{GI})\nonumber \\
&=&\left[C_{ii}^{gG}+g_iC_{ii}^{IG}\right]+\left[C_{ii}^{gI}+g_iC_{ii}^{GG}
\right]\ .
\ea
Here $g_i$ is the averaged $g(F)$ over galaxies in the $i$-th redshift
bin, $g_i=\langle g_i(F)\rangle$. $g(F)$ is of order unity
(e.g. \citealt{Scranton05}). However, since it changes sign from bright end 
of the luminosity function to the faint end, We expect that the averaged $g_i$
to  be smaller than $1$ for sufficiently deep surveys, $g_i<1$.

Our goal is to measure $C_{ii}^{gI}$, with new contaminations from the
magnification bias.  We can apply the same weighting of the estimator
Eq. \ref{eqn:estimator} here. On one hand, both $C_{ii}^{gI}$ and $C_{ii}^{GG}$
are unchanged by this weighting. On the other hand, both $C_{ii}^{GI}$
and $C_{ii}^{gG}$ are reduced by virtually the same $1-Q$. These behaviors
mean that, the estimator Eq. \ref{eqn:estimator} eliminates the combination
$C_{ii}^{gG}+g_iC_{ii}^{IG}$ and  measures the combination
$C_{ii}^{gI}+g_iC_{ii}^{GG}$ in which the term $g_iC_{ii}^{GG}$ contaminates
the I-g measurement. 

$g_iC_{ii}^{GG}$ can not be eliminated completely, due to various sources of
error. An obvious one is  the measurement error in $g(F)$. At bright end, the
galaxy
number density drops exponentially and lensing modifies $N(>F)$ significantly
for its steep slope. At faint end,  the flux measurement  noise is large.
Catastrophic
photo-z error is also an issue \citep{Schneider00}. We will not estimate these
errors for realistic surveys. Instead, we ask how stringent the requirement on
the $g_i$ and $C_{ii}^{GG}$ measurements should be in order to make the impact
of the magnification bias negligible.

 Suppose the $g_i$ measurement has an error $\delta g_i$ and the
$C_{ii}^{GG}$ measurement has an error $\delta C_{ii}^{GG}$,  the induced
fractional  error in $C_{ij}^{GG}$ measurement is 
\ba
 &\frac{W_{ij}\Delta_i}{b_i}\frac{\delta g_i C^{GG}_{ii}+g_i\delta
  C^{GG}_{ii}}{C^{GG}_{ij}}\no \\
<&\frac{W_{ij}\Delta_i}{b_i}\left(|\delta
g_i|+\left|g_i\frac{\delta C^{GG}_{ii}}{C^{GG}_{ii}}\right|\right)\no \\
=&O(10^{-3}) \left(\left|\frac{\delta
g_i}{0.1}\right|+\left|\frac{g_i\delta
  C^{GG}_{ii}/C^{GG}_{ii}}{10\%}\right|\right)\no\ .
\ea
The above relation holds since $C^{GG}_{ii}<C^{GG}_{ij}$ ($i<j$),
$b_i=O(1)$ and $W_{ij}\Delta_i=O(10^{-2})$.   (1) If $g_i$ itself is small ($|g_i|\la
0.1$), then there is no need to correct for the $g_iC^{GG}_{ii}$ term since
its influnece is at the level of $0.1\%$ and thus negligble. (2) If $g_i$ is
large, but it can be
measured with an accuracy $\pm 0.1$,  and if $C^{GG}_{ii}$ can be measured
with $10\%$ accuracy, the magnification bias induced error will be
$O(10^{-3})$. It can thus be 
safely neglected, comparing to the minimum statistical error in $C^{GG}_{ii}$
(Fig. \ref{fig:fij}) or to other residual errors of the self-calibration
technique (Fig. \ref{fig:fij} \& \ref{fig:epsilon}).  Direct measurement of $g_i$ from the observed (lensed) flux galaxy distribution  in the redshift bin
and the approximation $C^{(1)}_{ii}\simeq C^{GG}_{ii}$ likely meet the
requirement, unless the II contamination is larger than $10\%$.  (3) 
If the II contamination is larger than $10\%$ and the measurement of
$C^{GG}_{ii}$ is heavily polluted, a more complicated method may work.  We can
split galaxies into flux  bins and perform the above analysis to each flux 
bin. Since  $g(F)$ changes in a known way across these flux bins, we are in
principle able to eliminate the $g_iC_{ii}^{GG}$ term, combining all these
measurements.  For  example, one can find an appropriate weighting function $W(F)$
such that $\langle 
g(F)W(F)\rangle=0$. Although this method requires more accurate $g(F)$ 
measurement,  it does not require measurement on  $C^{GG}_{ii}$
and thus avoids the II contamination and other associated errors. 

Based on the above arguments,  we expect that our self-calibration technique
is safe  against the magnification bias, although extra care is indeed required. 

\subsection{Catastrophic photo-z error}
\label{subsec:photoz}
Numerical calculations we  perform in this paper only consider a Gaussian
photo-z PDF. Observationally, the photo-z PDF is more complicated, with
non-negligible outliers (e.g. \citealt{Jouvel09,Bernstein09b}). The existence
of this catastrophic error affects the self-calibration technique through two
ways. (1) It affects the accuracy of the $Q$ 
estimation. (2) It affects the scaling relation Eq. \ref{eqn:relation}.   As
shown in
  the appendix \S  \ref{sec:scaling} and further discussed in \S
  \ref{subsec:systematic}, a key condition in deriving 
  Eq. \ref{eqn:relation} is that the true galaxy distribution in a given photo-z
bin is sufficiently narrow and  smooth. So likely catastrophic error leads to
degradation of the scaling  relation 
(Eq. \ref{eqn:relation}).\footnote{However, some  forms of catastrophic error
  bring better match in the redshift evolution of 
  the integrands of Eq. \ref{eqn:IG} \& \ref{eqn:Ig} and thus can actually
  improve the accuracy of the scaling relation.} 

From the appendix B, catastrophic error  introduces bias to $Q$, mainly
through its impact on $\eta$. Stage IV lensing projects need to control the
outlier rate $f_{\rm
  cat}$ to $\sim 0.1\%$  accuracy \citep{Hearin10} in order for the induced
systematical errors to be sub-dominant. If it is the case, we are able to perturb
the photo-z PDF $p(z|z^P)$ in Eq. \ref{eqn:eta}. We choose $|z-z^P|>\Delta$ as
the criteria of the catastrophic error and then have   $f_{\rm cat}=\int_0^{z^P-\Delta}
p(z|z^P)dz+\int^{\infty}_{z^P+\Delta} p(z|z^P)dz$. Since $f_{\rm cat}\ll 1$, from
Eq. \ref{eqn:eta}, we  find that 
the induced bias $\delta Q=O(f_{\rm cat})$. As long as the goal $|f_{\rm
  cat}|\la 0.1\%$ can be achieved, the induced error in $Q$ is less than $1\%$
and hence not a significant source of error in the
self-calibration. Furthermore, we are able to infer the statistically averaged
photo-z PDF through self- and cross-calibrations of photo-z errors, even with
the presence of large catastrophic errors 
(e.g. \citealt{Schneider06,Newman08,Zhang09,Benjamin10}). Since $Q$ can be
predicted given the photo-z PDF,  we are able to reduce  the possible bias in
$Q$, even  if the actual $f_{\rm cat}\ga 0.1\%$. 

The catastrophic error also affects the scaling relation. It biases both
$C^{IG}$, through the term $W_j$ and $n_i$ in Eq. \ref{eqn:IG}, and
$C^{Ig}$, through the 
term $n_i^2$ in Eq. \ref{eqn:Ig}. Part of the effect cancels when taking the ratio of the
two. The residual error is also of the order $O(f_{\rm cat})$. Hence from the
above order of magnitude estimation, the bias induced by catastrophic error is
likely sub-dominant to the major systematical error $\delta f_{ij}^{(c)}$ in
the scaling relation. More sophisticated analysis is required to robustly quantify its
impact.


\subsection{Stochastic galaxy bias}
A key assumption in Eq. \ref{eqn:relation} is the deterministic galaxy
bias with respect to the matter distribution. In reality there is
stochasticity in galaxy distributions, which  can both cause random scatters
and systematic shift in the scaling relation. Quantifying its impact is beyond
our capability, since the  galaxy stochasticity, the intrinsic
alignment and correlation between the two are not well
understood. For example, the galaxy bias is likely correlated with the
strength of the intrinsic alignment, since both depend on the type of
galaxies. Nonetheless, there are hopes to control its impact. (1) The 
galaxy stochasticity can in principle be measured 
(e.g. \citealt{Pen98,Fan03,Bonoli08,Zhang08}) and modeled
(e.g. \citealt{Baldauf09}).  Measurement and modeling of the intrinsic
alignment can be improved too
(e.g. \citealt{Hirata04b,Okumura08,Schneider09}). (2) Recently
\citet{Baldauf09} showed that,
by proper weighting and modeling, the galaxy stochasticity can be suppressed
to $1\%$ level to $k\sim 1 h/$Mpc. Thus there is promise to
control the error induced by the galaxy stochasticity in the self-calibration
to be $\sim 1\%\times f_{ij}^I$. This error is sub-dominant to other
systematical errors,
especially the one induced by the scaling relation inaccuracy (\S
\ref{subsec:systematic}). 

\subsection{Cosmological uncertainties}
The self-calibration techniques require evaluation of $W_{ij}$ in
Eq. \ref{eqn:relation} and $Q$ in Eq. \ref{eqn:estimator}. Both evaluations
involve the cosmology-dependent lensing kernel $W_L(z_L,z_G)\propto 
\Omega_m(1+z_L)\chi_L(1-\chi_L/\chi_G)$.  Fortunately, we do not need strong
cosmological priors to evaluate it. $\Omega_m$ has already been measured to
$5\%$ accuracy \citep{Komatsu10}  and will be measured to below $1\%$ accuracy
by 
Planck.\footnote{http://www.rssd.esa.int/index.php?project=PLANCK\&page=perf\_top
} Stage III BAO and supernova surveys will measure the distance-redshift
relation to $1\%$ accuracy (e.g. \citealt{DETF}).  So if we take these priors,
uncertainties in
cosmology can at most bias the self-calibration at percent level accuracy,
negligible to  the identified
$\sim 10\%$  scaling relation error in  \S \ref{subsec:systematic}.  We need
further investigation to robustly quantify the impact of uncertainties in
cosmological parameters. 


\section{Discussions and Summary}
\label{sec:summary}
We have proposed a self-calibration technique to eliminate the GI
contamination in cosmic shear measurement. It contains two original
ingredients. (1) This technique is able to extract the I-g cross correlation
from  the galaxy density-ellipticity correlation of the same redshift bin in
the given lensing survey with photo-z measurement. (2) It
then converts this I-g measurement into a measure of the GI correlation
through a generic scaling relation. The self-calibration 
technique has only moderate requirement on the photo-z accuracy and results in
little loss of cosmological information. We 
have performed simple estimation on the performance of this self-calibration
technique, which  suggests that it can either render the systematical GI
contamination into a negligible statistical 
error, or suppress the GI contamination by a factor of $\sim 10$, whichever is
larger.  

The GI self-calibration can be combined  with the photo-z
self-calibration \citep{Zhang09} for a joint self-calibration against both the
GI contamination and the photo-z outliers. This combination does not over-use
the information in weak lensing surveys. The GI self-calibration mainly use
the galaxy ellipticity-density correlation in the same redshift bin. On the
other hand, the photo-z self-calibration mainly relies on the cross correlation
between
galaxy ellipticity-density correlation between different  redshift bins.

More robust and self-consistent evaluation of the self-calibration (GI and photo-z)
performance requires comprehensive analysis of all relevant errors discussed in
\S \ref{sec:error} \& \ref{sec:moreerrors}, and possibly more, along with
realistic model of galaxy bias and intrinsic alignment. We expect that our
self-calibration technique will still work under this more complicated and
more realistic situation,  since the method to extract the I-g correlation and
the scaling relation between I-g and I-G are robust against the complexities
mmentioned above.  Recently, \citet{Joachimi09b,Kirk10} proposed 
simultaneous fittings of cosmological parameters, galaxy bias and intrinsic
alignment. Our self-calibration
technique can be incorporated in a similar scheme.  

Our self-calibration technique only uses the shape-density and density-density
measurements in the {\it same} redshift bin to calibrate the intrinsic
alignment.  Lensing surveys contain more  information on the 
intrinsic alignment, in the shape-shape correlation of the {\it same} and
between
{\it different} redshift bins,  and shape-density correlation between {\it
  different} redshift bins.  These information has been incorporated by
\citet{Joachimi08,Joachimi09,Joachimi09b,Okumura09,Kirk10,Shi10}) to calibrate
the
intrinsic alignment.  These techniques are complementary to each other and
shall be combined together for optimal calibration.

\acknowledgments
{\it Acknowledgments}:  
The author thanks Yipeng Jing and Xiaohu Yang for useful information on galaxy
intrinsic ellipticity. The author thanks Gary Bernstein and the anonymous
referees for many useful suggestions and comments.
This work is supported   by the one-hundred-talents   
program of the Chinese academy of science (CAS), the national science
foundation of China (grant No. 10821302 \& 10973027) and the 973 program grant
No. 2007CB815401.

\appendix
\section[]{A: The scaling relation}
\label{sec:scaling}
We derive the scaling relation (Eq. \ref{eqn:relation}) under the
Limber approximation. Under this approximation, the 2D GI angular cross
correlation power spectrum between the $i$-th and $j$-th redshift bins is
related to the 3D matter-intrinsic alignment 
cross correlation power spectrum $\Delta^2_{mI}(k,z)$ by 
\ba
\label{eqn:IG}
\frac{\ell^2}{2\pi}C^{IG}_{ij}(\ell)=\frac{\pi}{\ell}\int_0^{\infty}
\Delta^2_{mI}\left(k=\frac{\ell}{\chi(z)},z\right)W_j(z)\chi(z) \bar{n}_i(z)dz \ .
\ea
Here, 
\be
\label{eqn:Wj}
W_j(z_L)\equiv \int_0^{\infty} W_L(z_L,z_G)\bar{n}_j(z_G)dz_G\ .
\ee
As a reminder, $\bar{n}_i(z)$ is the true redshift distribution of galaxies in
the $i$-th redshift bin and $W_L(z_L,z_G)$ is the lensing kernel. 
The integral limit runs from zero to infinite, to take into account the
photo-z errors.  On the other hand,  the 2D angular power spectrum between the
intrinsic alignment and galaxy number density in the $i$-th redshift bin is
\ba
\label{eqn:Ig}
\frac{\ell^2}{2\pi}C^{Ig}_{ii}(\ell)=\frac{\pi}{\ell}\int_0^{\infty}
\Delta^2_{gI}\left(k=\frac{\ell}{\chi(z)},z\right)n^2_i(z)\chi(z)\frac{dz}{d\chi}
dz=b_i(\ell) \frac{\pi}{\ell}\int_0^{\infty}
\Delta^2_{mI}\left(k=\frac{\ell}{\chi(z)},z\right)n^2_i(z)\chi(z)\frac{dz}{d\chi}
dz \ .
\ea
$\Delta^2_{gI}(k,z)$ is the 3D galaxy-galaxy intrinsic alignment  power
spectrum.  In the last relation we have adopted a deterministic galaxy bias
$b_g(k,z)$ with respect to matter distribution and thus
$\Delta^2_{gI}(k,z)=b_g(k,z)\Delta^2_{mI}(k,z)$. $b_i(\ell)$ is defined by the
above equation. It is the average of  $b_g(k=\ell/\chi,z)$ over the redshift
bin. As long as $b_g(k,z)$ does not  change
dramatically, we have $b_i(\ell)=b_g(k=\ell/\chi_i,\bar{z}_i)$, to a good
approximation.  

In the limit that the ture redshift distribution of
galaxies in the $i$-th redshift bin is narrow, $\Delta^2_{mI}$
($\Delta^2_{gI}$) changes slowly and can be approximated as
$\Delta^2_{mI}(k=\ell/\chi_i,\bar{z}_i)$
($\Delta^2_{gI}(k=\ell/\chi_i,\bar{z}_i$)). We then have the
following approximations, 
\ba
\label{eqn:IG2}
\frac{\ell^2}{2\pi}C^{IG}_{ij}(\ell)\simeq  \frac{\pi}{\ell}
\Delta^2_{mI}\left(\frac{\ell}{\chi_i},\bar{z}_i\right) W_{ij}\chi_i\ ,
\ea
and
\ba
\label{eqn:Ig2}
\frac{\ell^2}{2\pi}C^{Ig}_{ii}(\ell)\simeq
b_i(\ell) \frac{\pi}{\ell}
\Delta^2_{mI}\left(\frac{\ell}{\chi_i},\bar{z}_i\right)\frac{\chi_i}{
  \Delta_i}\ .
\ea
 The
quantity $W_{ij}$ and $\Delta_i$ are already defined by Eq. \ref{eqn:wij} \&
\ref{eqn:Deltai}. 
Based on the above two equations, we derive Eq. \ref{eqn:relation}, whose
accuracy is quantified in \S \ref{subsec:systematic}.

\section[]{B: evaluating the $Q$ parameter}
\label{sec:Q}
To derive the relation between $C_{ii}^{Gg}|_S$ and $C_{ii}^{Gg}$, namely,
$Q\equiv C_{ii}^{Gg}|_S/C_{ii}^{Gg}$, we will begin with
the real space angular correlation function. We denote the angular correlation
function between the shear at $z_G^P$ and galaxies at $z_g^P$ as
$w^{Gg}(\theta;z_G^P,z_g^P)$. Its average over the distribution of galaxies in
the $i$-th redshift bin is 
\ba
w^{Gg}_{ii}(\theta)&=&\int_{\bar{z}_i-\Delta z_i/2}^{\bar{z}_i+\Delta
z_i/2} dz_G^P\int_{\bar{z}_i-\Delta z_i/2}^{\bar{z}_i+\Delta
z_i/2} dz_g^P
w^{Gg}(\theta;z_G^P,z_g^P)n_i^P(z_G^P)n_i^P(z_g^P)dz_G^Pdz_g^P\nonumber \\
&=&\int_{\bar{z}_i-\Delta z_i/2}^{\bar{z}_i+\Delta z_i/2}
dz_G^P\int_{\bar{z}_i-\Delta z_i/2}^{\bar{z}_i+\Delta z_i/2} dz_g^P
\int_0^{\infty} dz_G\int_0^{\infty} dz_g
\left[w^{Gg}(\theta;z_G,
z_g)p(z_G|z_G^P)p(z_g|z_g^P)n_i^P(z_G^P)n_i^P(z_g^P)\right]
  \ .
\ea
Here, $p(z|z^P)$ is the photo-z PDF. 
As a reminder, we have normalized such that
\ba
\int_{\bar{z}_i-\Delta z_i/2}^{\bar{z}_i-\Delta z_i/2} dz^P
n_i^P(z^P)=\int_0^{\infty} n_i(z)dz=1\ . \no
\ea
Since 
\be
w^{Gg}(\theta;z_G,z_g)=\int \langle
\delta_m(\theta^{'};z_L)\delta_g(\theta^{'}+\theta;z_g)\rangle
W_L(z_L,z_G)dz_L\ ,
\ee 
where $\langle \cdots \rangle$ denotes the ensemble average and in practice denotes
equivalently the average over $\theta^{'}$ (the ergodicity assumption), we have
\ba
w^{Gg}_{ii}(\theta)&=&\int_{\bar{z}_i-\Delta z_i/2}^{\bar{z}_i+\Delta
z_i/2}
dz_G^P\int_{\bar{z}_i-\Delta z_i/2}^{\bar{z}_i+\Delta z_i/2} dz_g^P
\int_0^{\infty} dz_G\int_0^{\infty} dz_g \int_0^{\infty} dz_L \\
&&\times \left[\langle
\delta_m(\theta^{'};z_L)\delta_g(\theta^{'}+\theta;z_g)\rangle
W_L(z_L,z_G)p(z_G|z_G^P)p(z_g|z_g^P)n_i^P(z_G^P)n_i^P(z_g^P)\right]\nonumber \\
&=&\int_0^{\infty} dz_L\int_0^{\infty} dz_g \left[\langle
\delta_m(\theta^{'};z_L) \delta_g(\theta^{'}+\theta;z_g)\rangle
W_i(z_L) n_i(z_g)\right]\ .\nonumber
\ea
Notice that the lensing kernel $W_L(z_L,z_G)=0$ when $z_L\geq z_G$. The averge
over all pairs with $z_G^P<z_g^P$ gives the other correlation function, 
\ba
w^{Gg}_{ii}|_S(\theta)&=&\int \langle \delta_m(\theta^{'};z_L)
\delta_g(\theta^{'}+\theta;z_g)\rangle
W_i(z_L) n_i(z_g)dz_Ldz_g \eta(z_L,z_g)\ .
\ea
Here, 
\be
\label{eqn:eta}
\eta(z_L,z_g)=\frac{2\int_{\bar{z}_i-\Delta z_i/2}^{\bar{z}_i+\Delta
z_i/2}
dz_G^P\int_{\bar{z}_i-\Delta z_i/2}^{\bar{z}_i+\Delta z_i/2} dz_g^P
\int_0^{\infty} dz_G
  W_L(z_L,z_G)p(z_G|z_G^P)p(z_g|z_g^P)S(z_G^P,z_g^P)n_i^P(z_G^P)n_i^P(z_g^P)}{
\int_{\bar{z}_i-\Delta z_i/2}^{\bar{z}_i-\Delta z_i/2}
dz_G^P\int_{\bar{z}_i-\Delta z_i/2}^{\bar{z}_i+\Delta z_i/2} dz_g^P
\int_0^{\infty} dz_G 
  W_L(z_L,z_G)p(z_G|z_G^P)p(z_g|z_g^P)n_i^P(z_G^P)n_i^P(z_g^P)}\
\ ,
\ee
where the selection function $S(z_G^P,z_g^P)=1$ if $z_G^P<z_g^P$ and
$S(z_G^P,z_g^P)=0$
otherwise. The factor $2$ comes from the relation 
\be
\frac{\int_{\bar{z}_i-\Delta z_i/2}^{\bar{z}_i+\Delta z_i/2}
dz_G^P\int_{\bar{z}_i-\Delta z_i/2}^{\bar{z}_i+\Delta z_i/2} dz_g^P
p(z_G|z_G^P)p(z_g|z_g^P)n_i^P(z_G^P)n_i^P(z_g^P)}{\int_{\bar{z}_i-\Delta
\bar{z}_i/2}^{\bar{z}_i+\Delta z_i/2}
dz_G^P\int_{\bar{z}_i-\Delta z_i/2}^{\bar{z}_i+\Delta z_i/2} dz_g^P
p(z_G|z_G^P)p(z_g|z_g^P)S(z_G^P,z_g^P)n_i^P(z_G^P)n_i^P(z_g^P)}=2\ .
\ee

 The power spectra $C^{Gg}_{ii}$ and $C^{Gg}_{ii}|_S$ are the Fourier
 transform of $w^{Gg}_{ii}$ and  $w^{Gg}_{ii}|_S$, respectively. To evaluate
 these power  spectra, we again follow the Limber approximation, which states
 that the dominant  correlation signal comes from $z_L=z_g$.  We then  have  
\ba
\label{eqn:Q}
\frac{\ell^2C^{Gg}_{ii}(\ell)}{2\pi}&=&\frac{\pi}{\ell}\int_0^{\infty}
\Delta^2_{mg}\left(k=\frac{\ell}{\chi(z)},z\right)\chi(z)W_i(z)n_i(z)dz \ , \\ 
\frac{\ell^2C^{Gg}_{ii}|_S(\ell)}{2\pi}&=&\frac{\pi}{\ell}\int_0^{\infty}
\Delta^2_{mg}\left(k=\frac{\ell}{\chi(z)},z\right)W_i(z)\chi(z)n_i(z)\eta(z,z_g=z)dz \ .
\ea

The quantity that we want to evaluate is $Q(\ell)\equiv
C^{Gg}|_S(\ell)/C^{Gg}(\ell)$. Since it is the ratio of the two power spectra,
in which $\Delta^2_{mg}$, $W_i$ and $n_i$ roughly cancel,  to the first order,
$Q\simeq \eta$. The value of $\eta$ is determined by the relative contribution to $C^{Gg}$
from pairs with $z_G^P<z_g^P$ and pairs with $z_G^P>z_g^P$. If the two sets
have the same contribution, $\eta=1$ and $Q=1$. In the limit $\sigma_P\gg
\Delta z$, the contribution 
from the pair with $z_G^P<z_g^P$ to $C^{Gg}_{ii}$ approaches to that of the
pair with $z_G^P>z_g^P$. So we have  $\eta\rightarrow 1$ and $Q\rightarrow
1$. In this limiting case, we will be no  longer able to use this weighting
scheme to separate $C^{Gg}$ and $C^{Ig}$. On the other hand, if $\sigma_P\ll
\Delta z$, the pair with $z_G^P<z_g^P$ virtually does not contribute to
$C^{Gg}$, we will have $\eta\rightarrow 0$ and $Q\rightarrow 0$, as would
happen for spectroscopic redshifts.  As long as $Q$ deviates significantly
from unity, $C_{ii}^{Ig}$ can be separated from $C_{ii}^{Gg}$.   For a
LSST-like survey with $\Delta z=0.2$ and $\sigma_P=0.05(1+z)$, we numerically
evaluate $\eta(z,z_g)$ and $Q(l)$. We find that $Q\sim 1/2$
(Fig. \ref{fig:Q}). The significant deviation of $Q$ from unity is the
manifestation of relatively large photo-z error $\sigma_P$, across which the
lensing efficiency changes dramatically.

\section[]{C: the statistical error in extracting $C^{Ig}_{ii}$}
\label{sec:appendixC}
For the convenience, we will work on the pixel space to derive the statistical
error in extracting $C^{Ig}_{ii}$ from the galaxy shape (ellipticity)-density
measurement in the $i$-th photo-z bin. For a given redshift
bin, we first pixelize the data into sufficiently fine (and uniform) pixels of
photo-z and angular position. Each pixel, with label $\alpha$, has a 
corresponding photo-z $z^P_{\alpha}$ and corresponding angular position
$\vec{\theta}_{\alpha}$.   Each pixel
also has a measured overdensity 
$\delta_{\alpha}+\delta_{\alpha}^N$ and a measured ``shear'',
$\kappa_{\alpha}+I_{\alpha}+\kappa^N_{\alpha}$.  Here, the superscript ``N''
denotes the measurement noise, e.g., the shot noise. In total, there are $N_P$
pixels. Following the definition of  the angular power spectrum, we have
\ba
C^{(2)}(\ell)=N_P^{-2}\sum_{\alpha\beta}
\left[\delta_{\alpha}+\delta_{\alpha}^N\right]\left[\kappa_{\beta}+I_{\beta}+\kappa^N_{\beta}\right]\exp\left[i\vec{\ell}\cdot 
  (\vec{\theta}_{\alpha}-\vec{\theta}_{\beta})\right] \ , \nonumber
\ea
\ba
C^{(2)}(\ell)|_S=2N_P^{-2}\sum_{\alpha\beta}
\left[\delta_{\alpha}+\delta_{\alpha}^N\right]\left[\kappa_{\beta}+I_{\beta}+\kappa^N_{\beta}\right]\exp\left[i\vec{\ell}\cdot
  (\vec{\theta}_{\alpha}-\vec{\theta}_{\beta})\right]\times S_{\alpha\beta}
  \ . 
\ea
Here, $S_{\alpha\beta}=1$ when $z^P_{\alpha}>z^P_{\beta}$ and vanishes
otherwise. In the limit that $N_P\gg 1$, $\sum_{\alpha\beta}S_{\alpha\beta}=N^2_P/2$.
Namely, the average $\overline{S_{\alpha\beta}}=1/2$. 

The $C^{Ig}$ measurement error, from Eq. \ref{eqn:estimator}, is 
\ba
\delta C^{Ig}&=&\frac{1}{(1-Q)}N_P^{-2}\sum_{\alpha\beta}\exp\left[i\vec{\ell}\cdot 
  (\vec{\theta}_{\alpha}-\vec{\theta}_{\beta})\right]
\left[(\delta_{\alpha}+\delta_{\alpha}^N)(\kappa_{\beta}+I_{\beta}+\kappa^N_{\beta})(2S_{\alpha\beta}-Q)-(1-Q)\delta_{\alpha}I_{\beta}\right]\nonumber\\
&=&\frac{1}{(1-Q)}N_P^{-2}\sum_{\alpha\beta}\exp\left[i\vec{\ell}\cdot 
  (\vec{\theta}_{\alpha}-\vec{\theta}_{\beta})\right]
\left[(\delta_{\alpha}(\kappa_{\beta}+\kappa^N_{\beta})+\delta_{\alpha}^N(\kappa_{\beta}+I_{\beta}+\kappa^N_{\beta})\right](2S_{\alpha\beta}-Q)\ .
\ea
The last expression has utilized the relation $\overline{S_{\alpha\beta}}=1/2$
and the fact that the density-intrinsic alignment correlation does not depend
on the ordering along the line-of-sightof galaxy pairs. The rms error is 
\ba
(\Delta C^{Ig})^2&=&\frac{1}{(1-Q)^2}N_P^{-4}\sum_{\alpha\beta\rho\sigma}\exp\left[i\vec{\ell}\cdot 
  (\vec{\theta}_{\alpha}-\vec{\theta}_{\beta})\right] \exp\left[-i\vec{\ell}\cdot 
  (\vec{\theta}_{\rho}-\vec{\theta}_{\sigma})\right] (2S_{\alpha\beta}-Q)(2S_{\rho\sigma}-Q)\nonumber\\
&&\times \langle
\left[(\delta_{\alpha}(\kappa_{\beta}+\kappa^N_{\beta})+\delta_{\alpha}^N(\kappa_{\beta}+I_{\beta}+\kappa^N_{\beta})\right]
\left[(\delta_{\rho}(\kappa_{\sigma}+\kappa^N_{\sigma})+\delta_{\rho}^N(\kappa_{\sigma}+I_{\sigma}+\kappa^N_{\sigma})\right]
\rangle  \nonumber \ .
\ea
Here, $\langle \cdots \rangle$ denotes the ensemble average.  To proceed, we adopt
a common simplification in the lensing error analysis, namely the Wick theorem
for 4-point correlation (which holds strictly for Gaussian field), 
\ba
\langle ABCD\rangle=\langle AB\rangle \langle CD\rangle+\langle AC\rangle\langle
BD\rangle+\langle BC\rangle\langle AD\rangle\ ; \ A,B,C,D\in
\delta,\delta^N,\kappa,\kappa^N, I \ .\nonumber
\ea
Plug the above equation in and keep all non-vanishing terms, we then have 
\ba
\label{eqn:C3}
(\Delta C^{Ig})^2&=&\frac{1}{(1-Q)^2}N_P^{-4}\sum_{\alpha\beta\rho\sigma}\exp\left[i\vec{\ell}\cdot 
  (\vec{\theta}_{\alpha}-\vec{\theta}_{\beta})\right] \exp\left[-i\vec{\ell}\cdot 
  (\vec{\theta}_{\rho}-\vec{\theta}_{\sigma})\right] (2S_{\alpha\beta}-Q)(2S_{\rho\sigma}-Q)\nonumber \\
&&\times \left[\langle
\delta_{\alpha}\delta_{\rho}\rangle\langle
\kappa_{\beta}\kappa_{\sigma}\rangle+\langle
\delta_{\alpha}\delta_{\rho}\rangle\langle \kappa^N_{\beta}\kappa^N_{\sigma}\rangle+\langle \delta_{\alpha}^N\delta_{\rho}^N\rangle\langle
(\kappa_\beta+I_\beta)(\kappa_\sigma+I_\sigma)\rangle+ \langle
\delta_{\alpha}^N\delta_{\rho}^N \rangle\langle
\kappa^N_{\beta}\kappa^N_{\sigma})\rangle\right]\ .
\ea
Notice that the sum over  $\langle \delta_{\alpha}\kappa_{\beta}\rangle \langle
\delta_{\rho}\kappa_{\sigma}\rangle$ terms vanishes, resulting from the definition of
$Q$.  The sum over $ \langle \delta_{\alpha}\kappa_{\sigma}\rangle\langle
\delta_{\rho}\kappa_{\beta}\rangle$ terms also vanishes, due to the mis-match
between the  Fourier phases (e.g. the one $\propto
\vec{\theta}_\alpha+\vec{\theta}_\sigma$) 
and the angular dependence of the correlation functions
(e.g. $\langle \delta_{\alpha}\kappa_\sigma\rangle=w_{gG}(\vec{\theta}_\alpha-\vec{\theta}_\sigma)$) .  

Correlation functions in Eq. \ref{eqn:C3} depend on the absolute
pair separation, but not on the relative pair orientation. For example, 
$\langle  \delta_{\alpha}\delta_{\rho}\rangle=\langle
\delta_{\rho}\delta_{\alpha}\rangle=w_g(|\vec{\theta}_\alpha-\vec{\theta}_\rho|)$,
whose Fourier transform is the galaxy angular power spectrum $C^{gg}$. This
allows us to do the sums above  analytically and obtain the final expression
of the rms error  {\it for  a single $\ell$ mode}
\ba
\label{eqn:C4}
(\Delta
C^{Ig})^2&=&C^{gg}C^{GG}+\left[C^{gg}C^{GG,N}+C^{gg,N}(C^{GG}+C^{II})\right]\left[1+\frac{1}{3(1-Q)^2}\right]+C^{gg,N}C^{GG,N}\left[1+\frac{1}{(1-Q)^2}\right]
\ .
\ea
In the above expression, we have used the fact that noises only correlate at
zero-lag ($\langle \kappa_\beta^N\kappa_\sigma^N\rangle \propto
\delta_{\beta\sigma}$ and $\langle \delta_{\alpha}^N\delta_\rho^N\rangle
\propto \delta_{\beta\sigma}$) and   the following relations,
\ba
\frac{1}{N_P^4}\sum_{\alpha\beta\rho\sigma}(2S_{\alpha\beta}-Q)(2S_{\rho\sigma}-Q)&\simeq&(1-Q)^2\ ,\nonumber
\\
\frac{1}{N_P^3}\sum_{\alpha\beta\rho}(2S_{\alpha\beta}-Q)(2S_{\rho\beta}-Q)&\simeq&(1-Q)^2+\frac{1}{3}\nonumber\ ,\\
\frac{1}{N_P^2}\sum_{\alpha\beta}(2S_{\alpha\beta}-Q)^2&\simeq&(1-Q)^2+1\ .\nonumber 
\ea
In the above relations, we have neglected terms of the order $O(1/N_P)$ and
higher,  since the number of pixels $N_P\gg 1$. 

Each term in the r.h.s of Eq. \ref{eqn:C4} has specific physical meaning and
hence deserves brief explanation. 
\bi
\item  The first term  $C^{gg}C^{GG}$ is the cosmic
  variance arising from the  lensing and galaxy density fluctuations. The $Q$
  dependence drops out, since both $C^{(2)}$ and $C^{(2)}|_S$ sample the same 
cosmic volume and share the identical  (fractional) cosmic variance from this term.    

$C^{gg}C^{GG}$ is a familiar term in the cosmic variance of the ordinary
galaxy-galaxy lensing power spectrum.  However, the other familiar term, $C^{gG,2}$,
does not show up here.   This again is caused by the fact that
  both $C^{(2)}$ and $C^{(2)}|_S$ sample the same  cosmic volume and the
  cosmic variances inducing $C^{gG,2}$ cancel in the estimator Eq. \ref{eqn:estimator}. 
\item The last term $C^{gg,N}C^{GG,N}[1+1/(1-Q)^2]$ is the contribution from the
  shot noise in the galaxy distribution and random shape shot noise in shear
  measurement.  The $Q$ dependence can be understood as follows.  Such error
  in $C^{(2)}$ has two contributions, $\delta C_A$ from pairs with $z_g^P>z_G^P$ and
$\delta C_B$ from pairs with $z_g^P\leq z_G^P$. The 
total error is $(\delta C_A+\delta C_B)/2$. Since they come from different
pairs, these two errors are uncorrelated ($\langle \delta C_A\delta
C_B^*\rangle=0$), but they have the same 
dispersion $\langle |\delta C_A|^2\rangle=\langle |\delta
C_B|^2\rangle=2C^{gg,N}C^{GG,N}$.  
The factor $2$ here provides the correct rms noise in
$C^{(2)}$,  which is $C^{gg,N}C^{GG,N}$.  
Clearly the shot noise error in  $C^{(2)}|_S$ is $\delta C_A$.  Plug the above
relations into Eq. \ref{eqn:estimator}, we find that the shot noise
contribution is indeed the last term in Eq. \ref{eqn:C4}. Unlike the cosmic
variance term, which does not rely on $Q$, the shot noise term blows up when
$Q\rightarrow 1$.  This corresponds to the case that the photo-z error is too
large to provide any useful information and thus we are no longer able to separate
the Ig contribution form the Gg contribution. 
\item
The middle term  is the cross talk between cosmic variance and shot
noise. One can find similar  terms in  usual cross correlation statistical
error analysis.   
\ei
Interestingly, when $(1-Q)^2=1/3$ and when $C^{II}\ll C^{GG}$, Eq. \ref{eqn:C4}
reduces  to 
\ba
(\Delta
C^{Ig})^2&\simeq& (C^{gg}+2C^{gg,N})(C^{GG}+2C^{GG,N}), \ \ when\  Q\sim 1-\sqrt{1/3}=0.423
\ . 
\ea
This expression is identical to the usual expression of cross correlation
statistical error, expect for the factor $2$.

\end{document}